\documentclass[aps,a4paper, preprint, superscriptaddress,preprintnumbers,floatfix,nofootinbib,amsmath,amssymb]{revtex4}
\usepackage{url}
\usepackage{hyperref}
\usepackage{cleveref}
\usepackage{subfig}

\usepackage{amstext,amssymb}
\usepackage[section]{placeins}
\usepackage{amsmath}
\usepackage{graphicx}
\usepackage{xspace}
\usepackage{color}
\usepackage{xcolor}
\usepackage{float}
\usepackage{comment}
\usepackage{amstext,amssymb}
\usepackage{amsmath}
\usepackage{graphicx}
\usepackage[section]{placeins}
\usepackage{xspace}
\usepackage{color}
\usepackage{units}
\usepackage{array}
\usepackage[section]{placeins}
\usepackage{amsmath,bm}
\usepackage{amssymb}
\usepackage{gensymb}
\usepackage{soul}
\usepackage{textcomp}
\usepackage{slashed} 
\usepackage{multirow}
\usepackage{hyperref}
\usepackage{mathrsfs}
\usepackage{enumitem}
\usepackage{graphicx}
\usepackage{color}
\usepackage{comment}
\usepackage{epstopdf}
\usepackage{float}
\usepackage{units}
\usepackage{slashed}
\usepackage{amsmath,bm}
\DeclareUnicodeCharacter{2212}{-}
\usepackage{slashed} 
\usepackage{multirow}
\newcommand{\be}{\begin{equation}}
\newcommand{\ee}{\end{equation}}
\newcommand{\bea}{\begin{eqnarray}}
\newcommand{\eea}{\end{eqnarray}}
\newcommand{\nn}{\nonumber}

\def\s1{\hat s}

\usepackage{slashed}

\newcommand{\red}[1]{\textcolor{red}{#1}}

\newcommand{\nua}[1]{\ensuremath{\rlap{\kern-2.5pt\ensuremath{\overset{\scriptscriptstyle(-)}{\phantom{\nu}}}}{\ensuremath{{\nu}_{#1}}}}\xspace}

\begin{document}
\title{ Neutrino phenomenology, W mass anomaly \& muon $(g-2)$ in minimal type-III seesaw using $T^\prime$ modular symmetry}
\author{Priya Mishra }
\email{mishpriya99@gmail.com}
\affiliation{School of Physics,  University of Hyderabad, Hyderabad - 500046, India}
\author{Mitesh Kumar Behera}
\email{miteshbehera1304@gmail.com}
\affiliation{School of Physics,  University of Hyderabad, Hyderabad - 500046,  India}
%
\author{ Rukmani Mohanta}
\email{rmsp@uohyd.ac.in}
\affiliation{School of Physics,  University of Hyderabad, Hyderabad - 500046,  India}

\begin{abstract}
In this study, we attempt to introduce a model to illustrate neutrino phenomenology by incorporating  two right-handed fermion triplet superfields, i.e., $\Sigma_{R_j}$, in the presence of the modular symmetry $\Gamma_3^\prime \simeq A_4^\prime$, a double cover of the $A_4$ modular symmetry. The motivation in utilizing double cover is, so far only even modular forms were considered for constructing modular invariant models, but, in this case, it is possible to extend the modular invariance approach to general integral weight modular forms, i.e., the odd weight modular forms. Hence, this type of amalgamation between $T^\prime$ modular symmetry and minimally extending the seesaw can correctly explain the neutrino phenomenology. Additionally, we have made an attempt to accommodate the most recent measurement of the $W$ boson mass, published by the CDF-II collaboration and shed some light on the recent results of muon $(g-2)$. Finally, we have discussed lepton flavor violation in order to establish a constraint on the mass of right-handed fermion.

\end{abstract}

\maketitle
\flushbottom

\section{INTRODUCTION}
\label{sec:intro}
After the discovery of the Higgs boson, the Standard Model (SM) has gained widespread accomplishment. Numerous experiments have carefully scrutinized the SM predictions, proving it to be a successful theory of electroweak interactions  \cite{Workman:2022ynf}. Although the SM is exceptionally victorious in explaining the interactions up to the electroweak scale, it fails to elucidate mixing patterns in quark and lepton flavor sectors, mass hierarchies amid leptons and quarks, including the non-zero  neutrino masses. Hence, using symmetry consideration seems to be the most effective strategy. In support of the above, the non-abelian discrete flavor symmetry groups have helped us to understand the lepton mixing pattern, whose literature is quite extensive. Discrete flavour symmetries \cite{Altarelli:2010gt,Feruglio:2012cw,Bazzocchi:2009da,Pramanick:2019oxb,Das:2018rdf,Ding:2018tuj,Borah:2017dmk} combined with generalised CP symmetry \cite{Ding:2013hpa,Holthausen:2012dk,Feruglio:2013hia,Ding:2013bpa} can lead to fairly predicative models. Notably, the flavor symmetry group, which attempts the explanation of observed quark and lepton flavor mixing patterns, can also accommodate CP symmetry concurrently. For the illustration of non-zero neutrino mass within the roof of SM, a higher dimensional operator (i.e., dimension-5) was pioneered by Weinberg \cite{Weinberg:1979sa}. Due to certain drawbacks associated with higher dimensional operators, the alternate approach of introducing right-handed (RH) neutrinos became popular, leading to the seesaw mechanism. The exchange of heavy RH particles scales down the mass of neutrinos in a natural way. In support of the above, type-I \cite{Mohapatra:1979ia,bilenky2018introduction, branco2020type}, type-II \cite{Gu:2006wj,Luo:2007mq,Antusch:2004xy,Rodejohann:2004cg,Gu:2019ogb,McDonald:2007ka} and type-III \cite{Liao:2009nq,Ma:1998dn,Foot:1988aq,Dorsner:2006fx,Franceschini:2008pz,He:2009tf} seesaw models are based on the exchange of heavy right-handed $SU(2)_L$ singlet fermions, triplet scalars, and fermionic triplets respectively. While constructing the models theoretically using  discrete flavor symmetries, several flavon fields are required to keep the model invariant under the symmetry groups. These flavon fields also break the flavor symmetry group into different subgroups via their vacuum expectation value (VEV) acquisition, as seen in the neutrino and charged lepton sectors. This often complicates the model, as the leading order corrections are often subjected to the corrections from higher dimensional operators as a consequence of utilizing multiple flavon insertions. 

The above shortcomings can be pulled off by a recent, yet well-established modular invariance approach \cite{feruglioneutrino, Brax:1994kv, Dudas:1995eq, Binetruy:1995nt}. As an advantage, flavon fields are not needed anymore or minimized, and the symmetry breaking is performed by the VEV of complex modulus field $\tau$. Consequently, the  model can be constructed elegantly by using lesser flavon insertions. In the superpotential, higher dimensional operators  are governed exclusively by modular invariance. It is possible to produce highly predictive models for neutrino masses and mixing angles with modular flavor symmetry. The role of modular forms is played by dimensionless Yukawa couplings, which are functions of modulus $\tau$. Their transformation is governed by the Dedekind eta function instead of being constant in the case of the conventional discrete flavor symmetry approach. Moreover, quark and lepton fields have certain modular weights, which define the nontrivial transformation of these fields under modular forms. As a result, there is a myriad of literature available utilizing finite modular groups, i.e., $\Gamma_2 \simeq S_3$ \cite{Du:2020ylx,Mishra:2020gxg,Okada:2019xqk},  $\Gamma_3 (\Gamma_3^\prime) \simeq A_4 (A_4^\prime)$ \cite{Ding:2019zxk,Abbas:2020qzc,Ding:2021eva,Singh:2023jke, Mishra:2020fhy,Devi:2023vpe,Kashav:2022kpk,Kashav:2021zir,Charalampous:2021gmf,Chen:2020udk,Okada:2022kee,Benes:2022bbg, Liu:2019khw, King:2020qaj,Wang:2019xbo,Lu:2019vgm, Li:2021buv, Gogoi:2022jwf, Kobayashi:2019gtp, Nomura:2019xsb, Asaka:2019vev, Behera:2020sfe,Behera:2020lpd,Mishra:2022egy},  $\Gamma_4 \simeq S_4$ \cite{Penedo:2018nmg,Novichkov:2018ovf,Okada:2019lzv,deMedeirosVarzielas:2022ihu,deAnda:2023udh,King:2021fhl}, $\Gamma_5 \simeq A_5$ \cite{Novichkov:2018nkm} and $\Gamma_5^\prime \simeq A_5^\prime$ \cite{Wang:2020lxk,Behera:2021eut,Behera:2022wco,Wang:2021mkw}. While setting up the modular invariance approach, the modular weights considered in the assumption are mostly even. However, literature pertaining to the idea of double covering of $A_4$ symmetry, known as $T^\prime$ symmetry \cite{liu2019neutrino}, allows both even and odd modular weights for constructing the model.

The main highlight of this work is to accommodate the recent $W$-mass anomaly reported by  the CDF collaboration, i.e., $m_W^{CDF-II} = 80.4335 \pm 0.0094$ GeV \cite{CDF:2022hxs}, which establishes a deviation of $7\sigma$ from the SM prediction, i.e., $m_W^{SM}=80.357 \pm 0.006$ GeV \cite{ParticleDataGroup:2020ssz}. For the central values, the deviation is $\delta m_W = m_W^{CDF} - m_W^{SM} = 0.0765$ GeV, which is quite a fascinating result from the viewpoint of new physics. These observations  led to multiple discussions regarding its potential implications and interpretations, for instance, Zee model utilizing two Higgs doublets \cite{Chowdhury:2022moc}, scotogenic-Zee model \cite{ Dcruz:2022dao}, type-II Dirac seesaw by adding a vector-like a fermion and real scalar triplet \cite{Borah:2022obi}, utilizing singlet-doublet fermion \cite{Borah:2022zim},  and additionally with the MSSM \cite{Heinemeyer:2022ith}, in $U(1)_{(L_\mu - L_\tau)}$ model with vector-like leptons which mix with muon can solve this anomaly \cite{Baek:2022agi}, introduction of one isospin doublet vector-like lepton\cite{Nagao:2022oin}, singlet-triplet scotogenic dark matter model \cite{Batra:2022org}, vector-like quark models including the electroweak precision data \cite{Cao:2022mif}, hadronic contributions by performing electroweak fits \cite{Athron:2022qpo}, 
singlet scalar extensions of the SM in the context of the $W$-boson mass \cite{Sakurai:2022hwh}. In the type-III seesaw model, the additional inclusion of a light fermion singlet $N$ and a heavy scalar triplet has significant implications, as discussed in \cite{Ma:2022emu}, the scalar triplet is also utilized to explain $W$ mass \cite{Heeck:2022fvl,Popov:2022ldh,Kanemura:2022ahw}.

The organization of this paper is as follows. In Sec.\ref{sec:modular_sym}, we accentuate certain striking features of $T^\prime$ modular symmetry, while in Sec.\ref{sec:model_framework}, we discuss the model framework containing particles contributing towards expressing the superpotential for type-III seesaw and the associated mass matrices. Subsequently, in Sec.\ref{sec:numerical}, we accomplish the numerical analysis where a mutual parameter space is extracted, satisfying all the phenomena discussed in our model. In section\ref{sec:wboson}, we illustrate the $W$-mass anomaly from CDF-II results, and the recent results of muon $(g-2)$ are discussed in Sec.\ref{sec:g-2}. We have also discussed lepton flavor violating decay mode $ \mu \to e \gamma$ in Sec.\ref{sec:LFVs} for obtaining the constraint on the lightest heavy fermion mass $M_{R_1}$. Finally, in  Sec.\ref{sec:conclude}, we summarize our findings. 

\section{Modular Symmetry as Double cover}
\label{sec:modular_sym}
The modular group $\Gamma_N$ is a dimension two finite group (i.e., $2 \times 2$ matrices) with integer entries and  determinant being unity, also known as $SL(2, Z_N )$ or homogeneous finite modular group. One can establish the double cover group $\Gamma_N^\prime$ from $\Gamma_N$ by including another generator $R$ which is related to $-{\rm I} \in SL(2, Z)$ and commutes with all elements of the $SL(2, Z)$ group, such that the generators $S, ~T$ and $R$ of $\Gamma_N^\prime$ obey certain relations as given below:
\begin{equation}
    S^2=R,~(ST)^3=1,~T^N=1,~R^2=1~\mathrm{and}~RT = TR.
\end{equation}

\subsection{$\Gamma_3^\prime \simeq A_4^\prime$ modular symmetry}
Since  $N=3$, the dimension of the linear space defined by the computationally efficient mathematical deductions relating to $\Gamma(3)$ is $k+1$, with $k$ being the modular weight. As a result, dimension two is produced if we consider the lowest-order modular weight, $k=1$. Dedekind's eta function as expressed by eqn.(\ref{dedekind}) is defined in the upper half plane, i.e., $\mathcal{H}= \{ \tau \in \mathbb{C}~|~\rm{Im}(\tau) > 0 \}$, is what creates the modular space
\begin{equation}
    \eta(\tau) = q^{1/24} \prod \limits_{i=1}^\infty (1-q^n),~~~~ q \equiv e^{2\pi i \tau}.
    \label{dedekind}
\end{equation}
Also, the generators $S$ and $T$ transforms $\eta$ as
\begin{equation}
    \eta(\tau +1)=e^{\nicefrac{i\pi}{12}}~\eta(\tau), ~~~~~~ \eta(-\nicefrac{1}{\tau})= \sqrt{-i\tau}~\eta(\tau).
\end{equation}
As we are working in the linear space of $\Gamma(3)$, whose expression depending upon $\eta$ is given by~\cite{schultz2015notes}
\begin{equation}
\label{eq:Mk_Gamma3}\mathcal{M}_{k}(\Gamma(3))=\bigoplus_{a+b=k,\,a,b\ge0} \mathbb{C} \frac{\eta^{3a}(3\tau)\eta^{3b}(\tau /3 )}{\eta^k(\tau)}\,.
\end{equation}
As the dimension of $\mathcal{M}_{k}(\Gamma(3))$ is $k+1$, hence, for $k=1$ we can take the basis vectors to be
\begin{eqnarray}
\nonumber&& \hat{e}_1(\tau)=\frac{\eta^{3}(3\tau)}{\eta(\tau)},\quad \hat{e}_{2}(\tau)=\frac{\eta^{3}(\tau / 3)}{\eta(\tau)}\,.
\end{eqnarray}
The basis vectors shown above are linearly independent, and any modular forms of $k=1$ and $N=3$ can be expressed as a linear combination of $\hat{e}_1$ and $\hat{e}_2$.
Further, due to application of generator $T$, $\hat{e}_i$ ($i=1, 2$) transform as
\begin{eqnarray}
\hat{e}_1(\tau)\stackrel{T}{\longmapsto} e^{i2\pi /3}\hat{e}_1(\tau),\qquad \hat{e}_{2}(\tau)\stackrel{T}{\longmapsto} 3(1-e^{i2\pi /3})\hat{e}_1 + \hat{e}_2 \,.
\end{eqnarray}
Similarly, under generator $S$
\begin{eqnarray}
\hat{e}_1(\tau) \stackrel{S}{\longmapsto} 3^{-3/2}(-i\tau)\hat{e}_2(\tau),\qquad \hat{e}_{2}(\tau)\stackrel{S}{\longmapsto} 3^{3/2}(-i\tau)\hat{e}_1(\tau) \,.
\end{eqnarray}
Therefore, utilizing the above information, one will be able to construct a modular multiplet $Y^{(1)}_{\mathbf{2}}$ which transforms as a doublet $\mathbf{2}$ under $\Gamma'_3 \cong T^\prime$ involving the basis vectors $\hat{e}_1$ and $\hat{e}_2$:
\begin{equation}
\label{eq:modular_space}
Y^{(1)}_{\mathbf{2}}(\tau)=\begin{pmatrix}
Y_1(\tau) \\
Y_2(\tau)
\end{pmatrix}\,,
\end{equation}
with
\begin{equation}
Y_1(\tau)=\sqrt{2}\,e^{i 7\pi/12}\,\hat{e}_1(\tau),\qquad Y_2(\tau)=\hat{e}_1(\tau)-\frac{1}{3}\hat{e}_2(\tau)\,.
\end{equation}
Further the higher weight modular Yukawa couplings with $k=2,3,4,5$ can be constructed from the tensor product of $Y^{(1)}_{\mathbf{2}}$ (see ref.\cite{liu2019neutrino}). Also the complete form of other doublet Yukawa couplings are mentioned in appendix \ref{app:A}. Refs \cite{Novichkov:2020eep, Liu:2020akv, Kikuchi:2020nxn} also discuss the double covering of group $\Gamma_N$.
\section{Model Framework }
\label{sec:model_framework}
To incorporate minimal type-III seesaw in our model, we have added right-handed hyperchargeless ($Y=0$) fermionic triplet superfields $\Sigma^c_{R_j}$ $( j=1,2)$, which transform as triplet under $SU(2)_L$ and  doublet under $T^\prime$ modular symmetry with $k_I = 3$. Further, Higgs  super-multiplets $H_{u,d}~ (Y= \pm 1/2)$ are singlets under $T^\prime$ modular symmetry with zero modular weight. The VEVs of Higgs super-multiplets i.e., $(v_u, v_d)$ are related to the SM Higgs VEV $(v_{\scriptscriptstyle H})$ by a simple equation $v_{\scriptscriptstyle H} =\frac12 \sqrt{v_u^2 + v_d^2}$. The ratio of Higgs super-multiplets VEVs is written as $\tan\beta = ({v_u}/{v_d}) \simeq 5$ (used in our analysis) \cite{Antusch:2013jca, Okada:2019uoy, Bjorkeroth:2015ora}. The SM right handed charged leptons $E_{1R}^c$, $E_{2R}^c$ and $E_{3R}^c$ transform as $1,1^\prime,$ and $ 1^{\prime\prime}$ under $T^\prime$ modular symmetry with $k_I = -2$. While, left handed (LH) lepton doublets $l_{Li} (i= e,\mu,\tau)$ transform as $1,  1^{\prime\prime}$ and $1^\prime$ under $T^\prime$ symmetry respectively with $k_I = 2$ represented in Table \ref{tab:fields-linear}.
\begin{table}[htpb]
\begin{center}
\begin{tabular}{|c||c|c|c|c|c||c|} \hline 
Fields & ~$E^c_{1R}$~& ~$E^c_{2R}$~  & ~$E^c_{3R}$~& ~${l_{L_i}}$~& ~$\Sigma^c_{R}$~&~ $H_{u,d}$\\ \hline \hline
$SU(2)_L$~&~$1$~&~$1$~&~$1$~&~$2$~&~$3$~&~$2$\\ \hline
$U(1)_Y$~&~$1$~&~$1$~&~$1$~&~$-\frac{1}{2}$~&~$0$~&~$\frac{1}{2}$,$-\frac{1}{2}$~\\ \hline
$T^\prime$~&~$1$~&~$1^\prime$~&~$1^{\prime\prime}$~&~$1,1^{\prime\prime},1^{\prime}$~&~$2$~&~$1$\\ \hline
$k_I$ & $-2$ & $-2$ & $-2$ & $2$ & $3$ & $0$ \\ \hline
\end{tabular}
\caption{Particle content of the model and their charges under $SU(2)_L \times U(1)_Y \times T^\prime$ group and their modular weights $k_I$.}
\label{tab:fields-linear}
\end{center}
\end{table}
\begin{table}[h!]
\centering
  \resizebox{\textwidth}{!}{  
\begin{tabular}{|c|c|c|c|c|c|}\hline  
Couplings~ & \multicolumn{1}{c||}{$Y_{2,I}^{(5)} = (y_{12},y_{22})$} & \multicolumn{1}{c||}{$Y_{2^\prime,I}^{(5)}$ = $(y_{12^\prime},y_{22^\prime})$}& \multicolumn{1}{c||}{$Y_{2^{\prime\prime},I}^{(5)} = (y_{12^{\prime\prime}},y_{22^{\prime\prime}})$} & \multicolumn{1}{c|}{$\lambda_1= Y_{3,I}^{(6)} = {(y_{13},y_{23},y_{33})}$}  & \multicolumn{1}{c|}{$\lambda_2=Y_{2^{\prime\prime}}^{(3)}$} \\ \hline \hline 
$T^\prime$ & ~$2$~& ~$2^\prime$~& ~$2^{\prime\prime}$~&~$3$~&~$2^{\prime\prime}$~\\ \hline
$k_I$ & ~$5$~& ~$5$~&~$5$~&~$6$~&~$3$~ \\ \hline
\end{tabular}
}
\caption{Charge assignment to Yukawa couplings under $T^\prime$ and its modular weight $k_I$.}
\label{tab:Yukawa}
\end{table}\\\\
\\
The complete superpotential is given by

\begin{eqnarray}
 \mathcal{W}  
                   &=&  \sqrt{2}y_{{\ell}} l_{L_i} H_d E^c_{R_i} +\alpha_D \left[Y_{\wp}^{(5)} H_u^T \eta (\Sigma^c_{R_j} {l_{L_i})_{\wp^\prime}}\right]+ \frac{M_{\Sigma} \alpha_\Sigma}{2}   {\rm Tr}\left[\sum \limits_{j=1}^2 \ {\Sigma_{R_j}^c} \lambda_1 \Sigma_{R_j}^c\right] \nonumber \\ &+& \mu H_u H_d + \lambda_1 M_{\Tilde{\Sigma}} {\rm Tr}[\Tilde{\Sigma}_j \Tilde{\Sigma}_j] +  \lambda_2 \left[ H_u^T \eta \Tilde{\Sigma}_1 H_d \right],
                   \label{comp_superpotential}
\end{eqnarray}
where, $\wp = (\textbf{2}^{\prime \prime},\textbf{2},\textbf{2}^{\prime})$, $\wp^\prime = (\textbf{2},\textbf{2}^{\prime \prime},\textbf{2}^{\prime}) $ with $\alpha_{\Sigma(D)} $ and $\Sigma^c_{R_j}$ are defined as,
\begin{eqnarray}
 \Sigma^c_{R_j}= \frac{1}{\sqrt{2}}
\begin{pmatrix}
\Sigma_j^{0c} & \sqrt{2}\Sigma_j^{+c} \\
\sqrt{2}\Sigma_j^{-c} & -\Sigma_j^{0c}
\end{pmatrix},~~~
\alpha_{\Sigma (D)}= \begin{pmatrix}
g_{\scriptscriptstyle \Sigma_1(D_1)} & 0\\
0 & g_{\scriptscriptstyle \Sigma_2(D_2)}
\end{pmatrix},~~~
\eta = \begin{pmatrix}
0&1\\
-1&0
\end{pmatrix},
\label{free-para}
\end{eqnarray}
with $\alpha_{\Sigma(D)}$ being the free parameter matrices, whereas, $M_\Sigma$ is the free mass parameter  and $\Tilde{\Sigma}_j$ being the scalar super-partner of triplet superfield ($\Sigma_j$). Further, $\lambda_1$ and $\lambda_2$ are the couplings with modular forms given  in Appendix \ref{app:A}. Table \ref{tab:Yukawa} contains the modular weights of the Yukawa couplings $(Y_{\wp}^{(5)})$ with $(\wp = (\textbf{2},\textbf{2}^\prime,\textbf{2}^{\prime \prime}))$, $\lambda_1$ and $\lambda_2$ along with their transformation under $T^\prime$ symmetry. Moreover, the charged lepton superpotential term is shown by the first part in eqn.(\ref{comp_superpotential}) yields a mass matrix (i.e., diagonal) exactly of the form as elaborated in ref. \cite{Mishra:2022egy}. Hence, we focus on the neutral lepton sector, as discussed below.\\\\
\underline{\textbf{Dirac mass term}}\\\\
The Dirac mass matrix for the neutral lepton sector can be  obtained from the following superpotential term:
 \begin{equation}
    \mathcal{W}_D = \alpha_D \sqrt{2} \left[Y_{\textbf{2}^{\prime \prime},\textbf{2},\textbf{2}^{\prime}}^{(5)} H_u^T \eta (\Sigma^c_{R_j} {l_{L_i})_{\textbf{2},\textbf{2}^{\prime\prime},\textbf{2}^\prime}} \right].
 \end{equation}
As $H_u$ gains the VEV, the neutral leptons obtain their masses. To make the Dirac term invariant, fermion triplets transform as doublet under $T^\prime$ modular symmetry. Hence, the Dirac interaction term of neutral multiplet of fermion triplet with the SM left-handed neutral leptons can be written as:
\begin{align}
M_D&=v_u
\left[\begin{array}{ccc}
y_{22^{\prime\prime}} & -y_{22} & -y_{22^\prime}  \\
-y_{12^{\prime\prime}} & y_{12} & y_{12^\prime}
\end{array}\right].    
\end{align}
\underline{\textbf{Majorana mass term}}\\\\
The superpotential for Majorana mass term for right-handed
neutrinos is given as
\begin{equation}
 \mathcal{W}_R = \frac{\alpha_\Sigma M_{\Sigma} }{2}~   {\rm Tr}\left[\sum \limits_{j=1}^2 \ {\Sigma_{R_j}^c} \lambda_1 \Sigma_{R_j}^c\right],
\end{equation}
where, $M_\Sigma$ is the free mass parameter and application of the $A_4^\prime$ product rule yields the mass structure as given below
\begin{align}
M_{R}&= \frac{M_\Sigma }{\sqrt2}\left[\begin{array}{cc}
g_{\Sigma_1} & 0  \\ 
0 & g_{\Sigma_2}   \\ 
\end{array}\right]
\left[\begin{array}{cc}
 \sqrt{2} e^{5\pi i/12}y_{23} & -y_{33} \\
 -y_{33} &\sqrt{2} e^{7\pi i/12}y_{13} \\
\end{array}\right] .
\end{align}
Thus, the active neutrino mass matrix in the framework of the type-III seesaw is given as
\begin{align}
m_\nu&= -M_D^T M_R^{-1} M_D.
\label{nmass}
\end{align}

\section{Numerical Analysis}
\label{sec:numerical}
The neutrino oscillation data from the NuFIT \cite{Esteban:2020cvm,website}  within their $3\sigma$ range serves as the reference for the numerical analysis for our model framework, as given in Table \ref{table:expt_value}.
\begin{table}[htbp]
\centering
\begin{tabular}{|c|c|c|c|c|c|c|}
\hline
\bf{Oscillation Parameters} & \bf{Best fit value} \bf{$\pm$ $1\sigma$} &  \bf{3$\sigma$ range} \\
\hline \hline
$\Delta m^2_{21}[10^{-5}~{\rm eV}^2]$ & $7.41^{+0.21}_{-0.20}$ & 6.82--8.03  \\
\hline
$|\Delta m^2_{31}|[10^{-3}~{\rm eV}^2]$ (NO) &  $2.507^{+0.026}_{-0.027}$ &  2.427--2.59\\
\hline
$\sin^2\theta_{12}$  & $0.303^{+0.012}_{-0.012}$ &  0.27--0.341\\
\hline
$\sin^2\theta_{23} $ (NO)
	  &	0.451$^{+0.019}_{-0.016}$ & 0.408--0.603 \\
\hline
$\sin^2\theta_{13} $ (NO) & 0.02225$^{+0.00056}_{-0.00059}$ & $0.02052-0.02398$ \\

\hline 
$\delta_{CP} /^\circ$ (NO) & 232$^{+36}_{-26}$ & $144-350$\\
\hline
\end{tabular}
\caption{The NuFIT values of the  oscillation parameters along with their 1$\sigma$/3$\sigma$ ranges.}
\label{table:expt_value}
\end{table}
The neutrino mass formula presented in eqn.(\ref{nmass}) leads for the deduction of the associated mass matrix on which numerical diagonalization is performed using the relation $\mathcal{U}^\dagger \mathcal{M} \mathcal{U}= {\rm diag}(m_{\nu_1}^2, m_{\nu_2}^2, m_{\nu_3}^2)$, where, ${\cal M}=m_\nu m_\nu^\dagger$ and $\mathcal{U}$ is the unitary matrix, from which the neutrino mixing angles can be derived using the conventional relations:
\begin{eqnarray}
\sin^2 \theta_{13}= |\mathcal{U}_{13}|^2,~~~~\sin^2 \theta_{12}= \frac{|\mathcal{U}_{12}|^2}{1-|\mathcal{U}_{13}|^2}\;,~~~~~\sin^2 \theta_{23}= \frac{|\mathcal{U}_{23}|^2}{1-|\mathcal{U}_{13}|^2}\;.
\end{eqnarray}
Another intriguing observable related  to the mixing angles and phases of the PMNS matrix is the Jarlskog invariant, expressed as
\begin{eqnarray}
J_{CP} &=& \text{Im} [\mathcal{U}_{e1} \mathcal{U}_{\mu 2} \mathcal{U}_{e 2}^* \mathcal{U}_{\mu 1}^*] = s_{23} c_{23} s_{12} c_{12} s_{13} c^2_{13} \sin \delta_{CP}\;.
\end{eqnarray}
Further, we chose the following model parameter ranges to fit the present neutrino oscillation data:
\begin{eqnarray}
&&{\rm Re}[\tau] \in [-0.5,0.5],~~{\rm Im}[\tau]\in [0.75,2],
\quad M_\Sigma \in  [10^4,10^5] \ {\rm TeV} , \nn\\
&&~~~~~~~~~~~~~~~~~~~\alpha_D \in  [10^{-5},10], ~~~~\alpha_{\Sigma} \in  [{10^{-2}},10^{-1}]\;.
\label{model-ranges}
\end{eqnarray}
We consider the free mass parameter ($M_\Sigma$), real and imaginary parts of $\tau$, free parameters $\alpha_D$ and $\alpha_\Sigma$ to vary randomly in their corresponding ranges\footnote{It is to be noted here that,  as seven free parameters (i.e., 2D matrices$-(\alpha_{\Sigma}, \alpha_{D})$, Re$(\tau)$, Im$(\tau)$, $M_\Sigma$) are being varied randomly to illustrate  the observed oscillation data by imposing certain constraint conditions, hence, the obtained correlations between different measured parameters  are less prominent.} given in eqn.(\ref{model-ranges}). 
The ranges for $\tau$'s real and imaginary parts are varied within $[-0.5,0.5]$ and $[0.75,2]$, respectively.
We noticed  that the model satisfies normal hierarchy (NH) scheme. 
We arbitrarily examine the parameter input values based on these ranges, hence, able to simultaneously satisfy the constraints on the sum of neutrino   masses obtained of Planck data \cite{Planck:2018vyg, Vagnozzi:2017ovm}, in the context of  the present model framework.

\begin{figure}[htpb]
\begin{center}
\includegraphics[height=50mm,width=75mm]{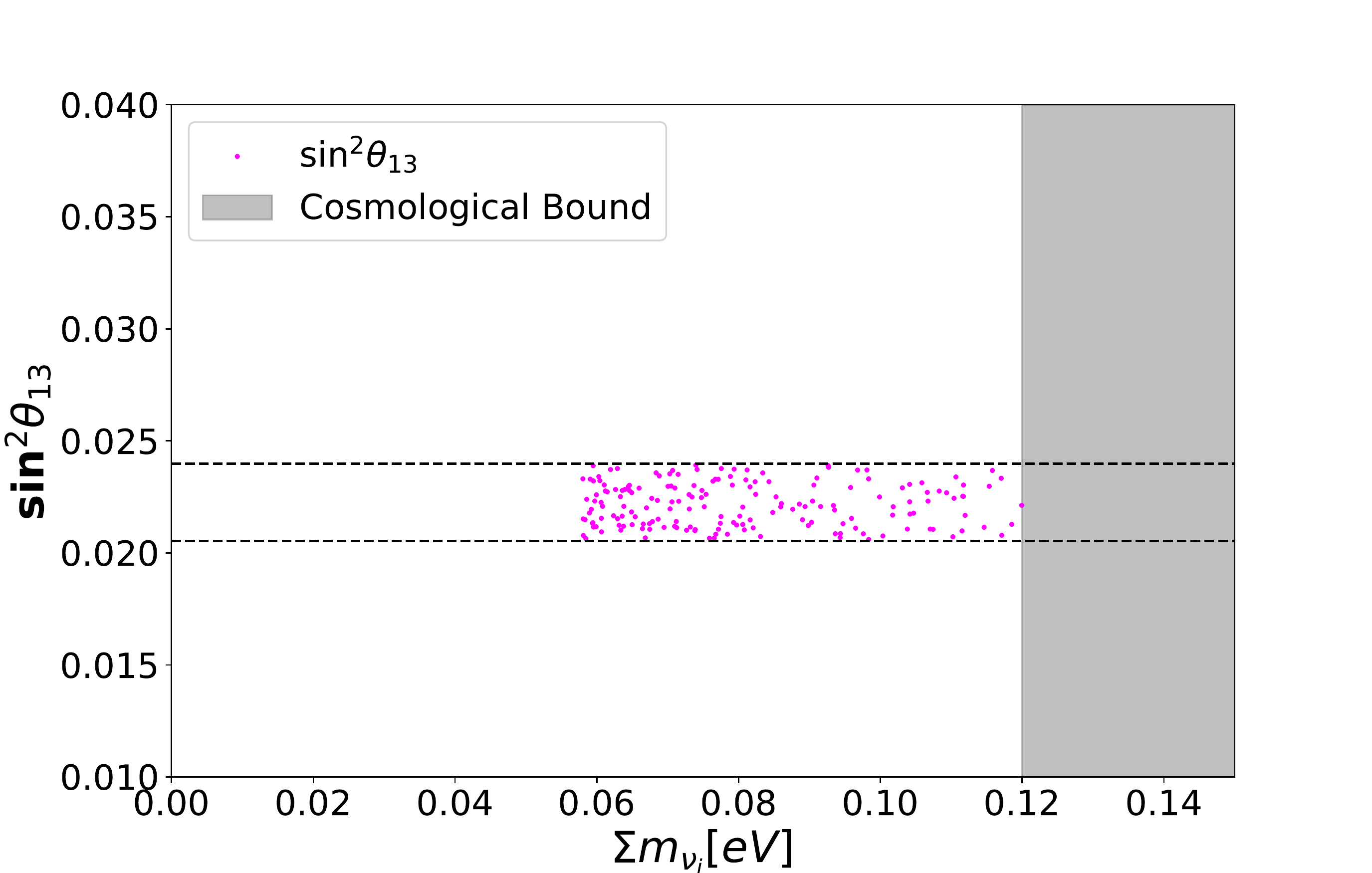}
\hspace*{0.2 true cm}
\includegraphics[height=50mm,width=75mm]{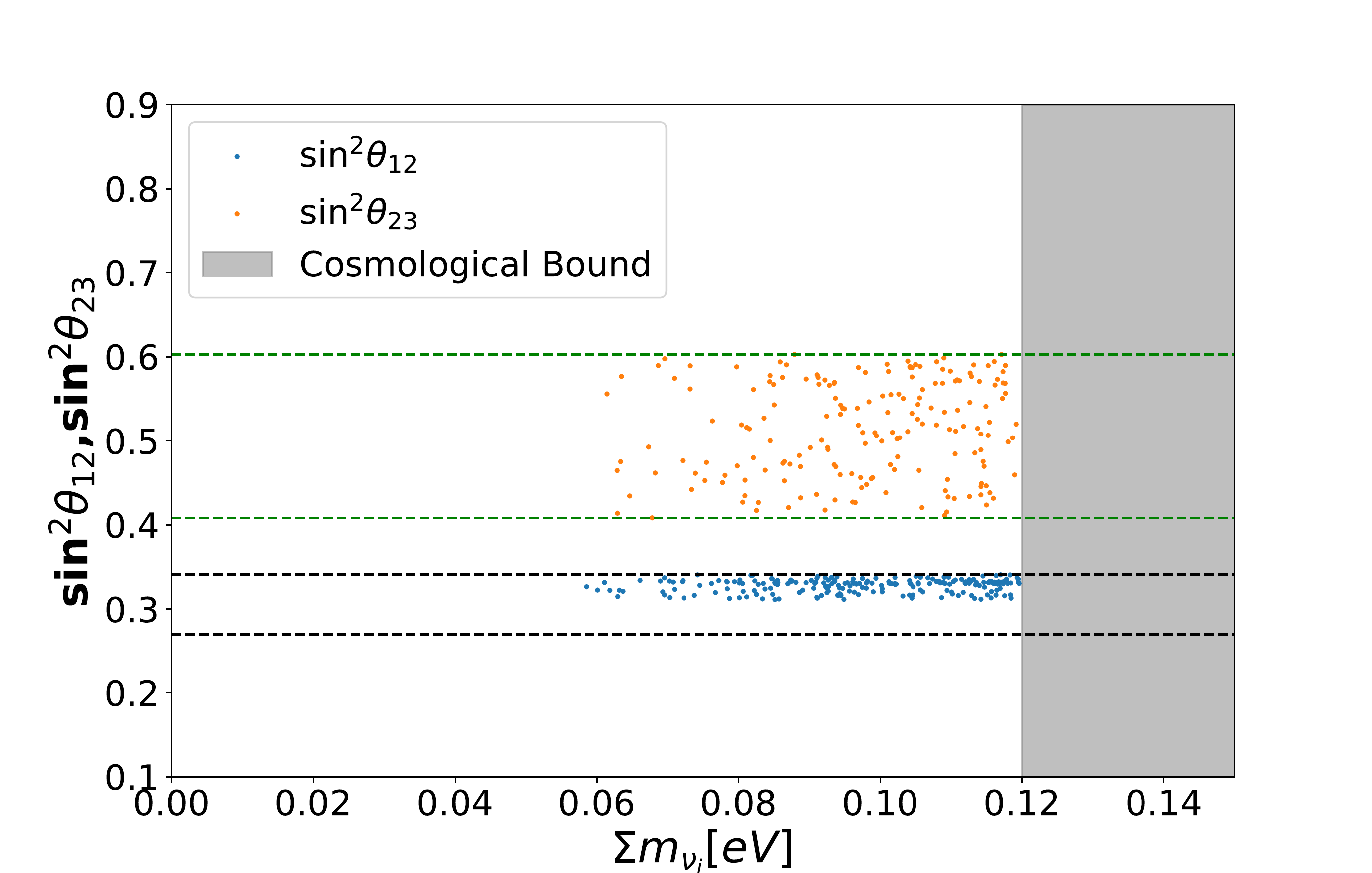}
\caption{Left (right) panel shows the  plane of  the mixing angles, i.e., $\sin^2 \theta_{13}$~($\sin^2 \theta_{12}$ \& $\sin^2 \theta_{23}$) with the sum of neutrino masses for the aforementioned ranges of model parameters while horizontal gridlines represent the $3\sigma$ range of mixing angles with the grey band being the excluded region from the cosmological bound (i.e., $\sum m_i \geq 0.12$ eV).}
\label{correl_mix_sum}
\end{center}
\end{figure}

\begin{figure}[h!]
\begin{center}
\subfloat[]{\includegraphics[height=50mm,width=58mm]{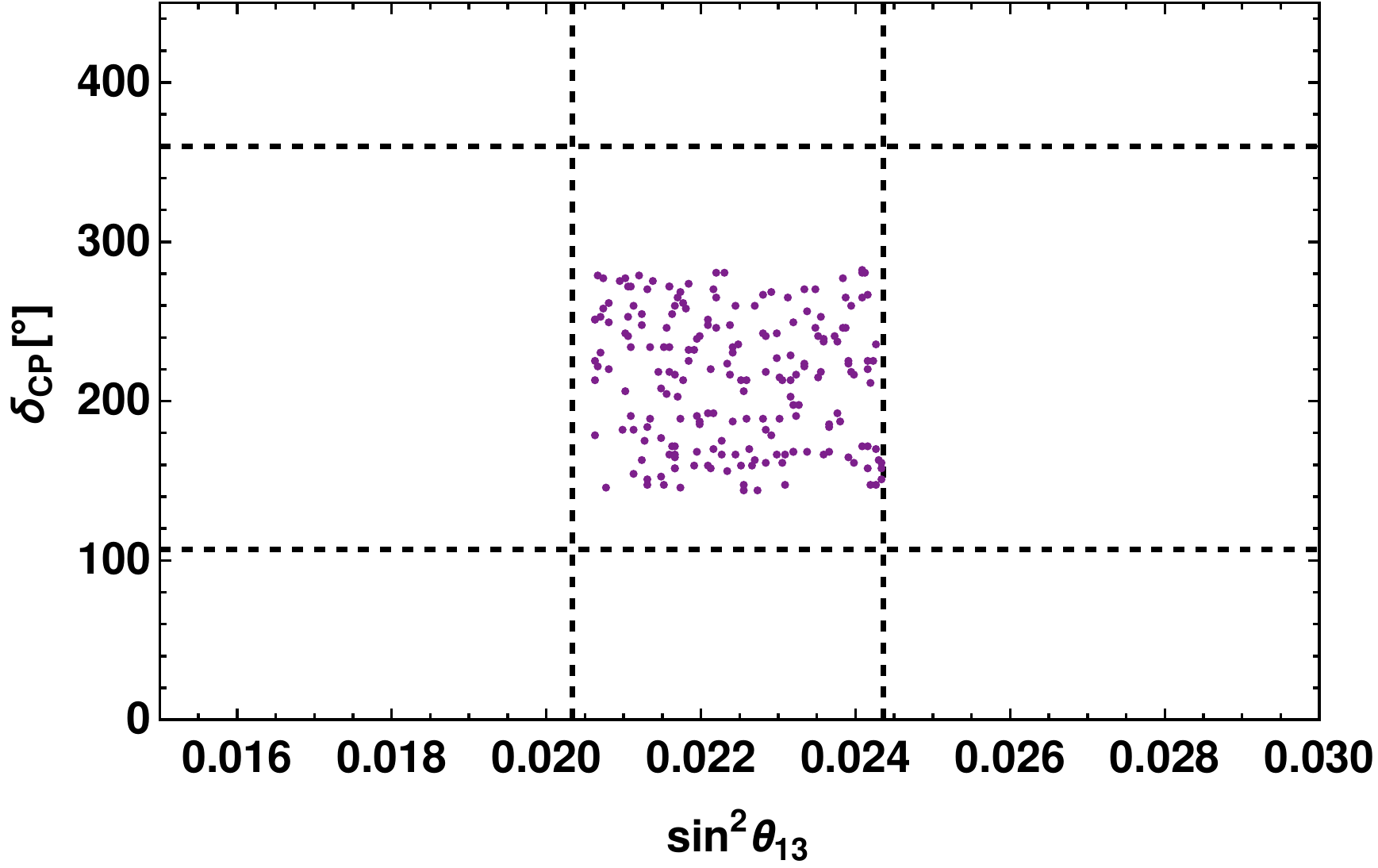}\label{dcp-ss13}}
\subfloat[]{\includegraphics[height=50mm,width=58mm]{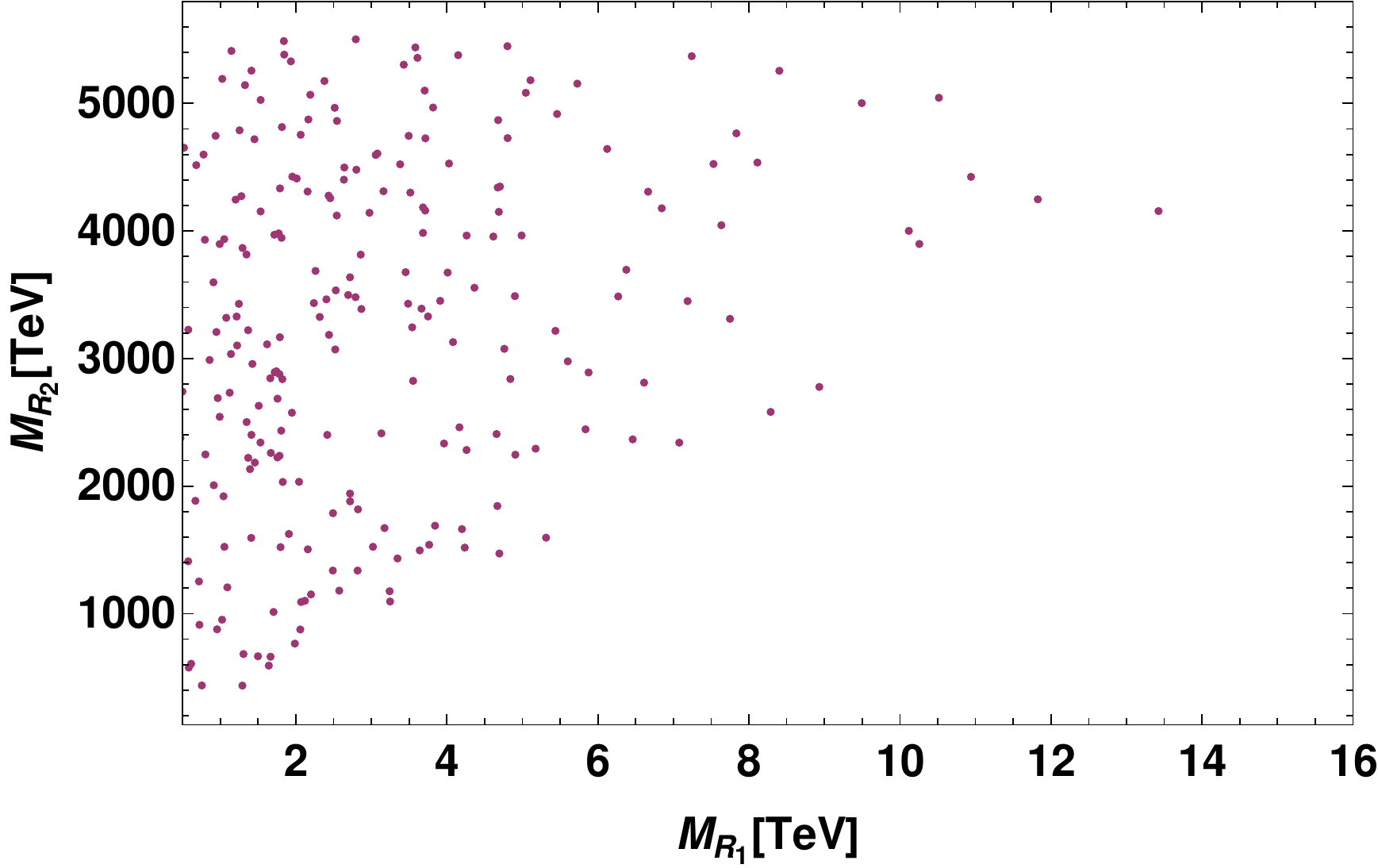}\label{M1-M2}}
\subfloat[]{\includegraphics[height=50mm,width=58mm]{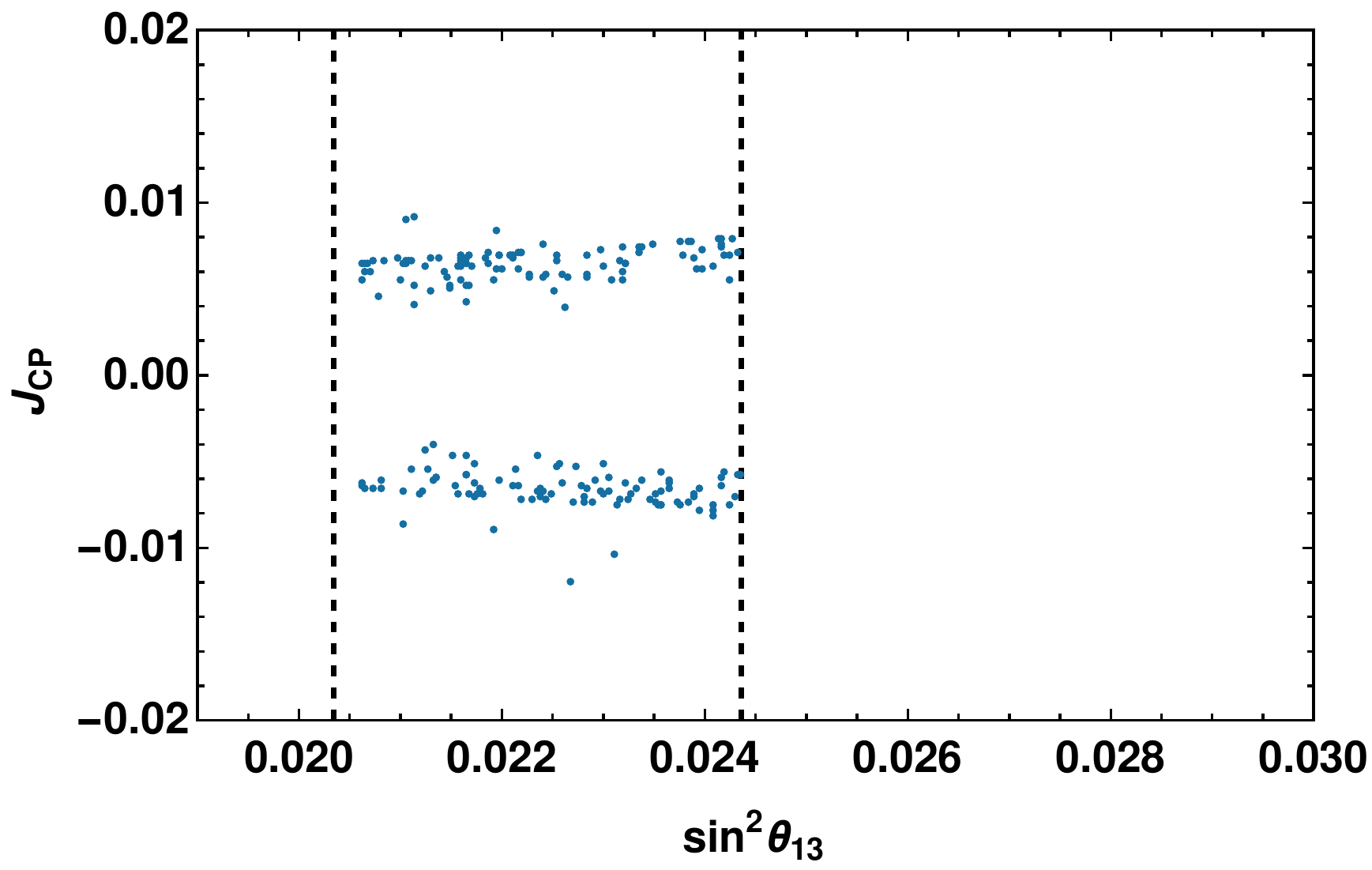}\label{jcp-ss13}}
\caption{Left (right) panel expresses the correlation between $\delta_{CP}$ ($J_{CP}$) w.r.t. mixing angle $\sin^2 \theta_{13}$, whereas, the middle panel depicts the correlation between heavy neutrino mass $M_{R_1}$ and $M_{R_2}$ in TeV scale.}
\label{obs} 
\end{center}
\end{figure}

\begin{figure}
    \centering
    \subfloat[]{\includegraphics[width=75mm, height=50mm]{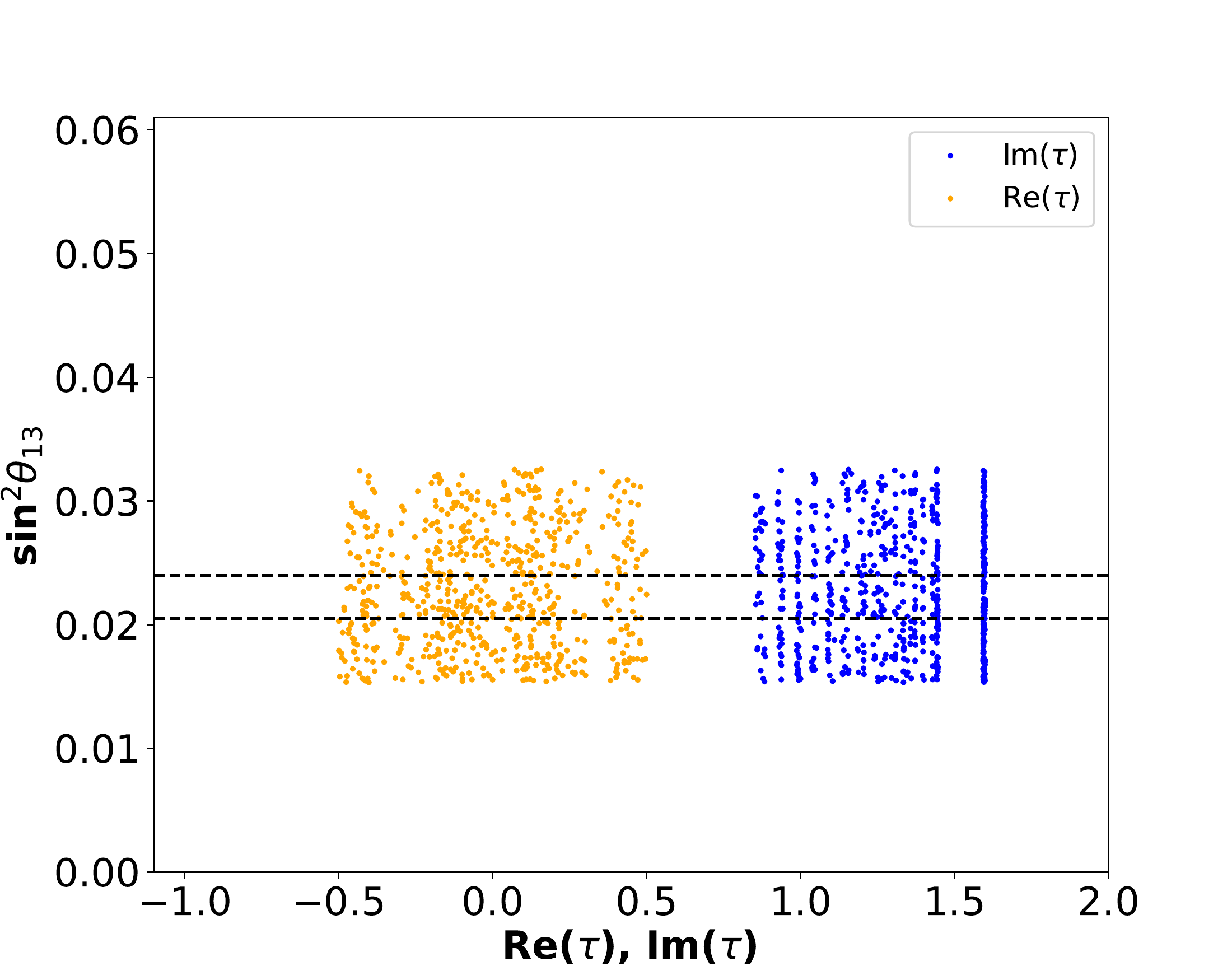}\label{rel-1}}
\hspace*{0.2 true cm}
    \subfloat[]{\includegraphics[width=75mm, height=50mm]{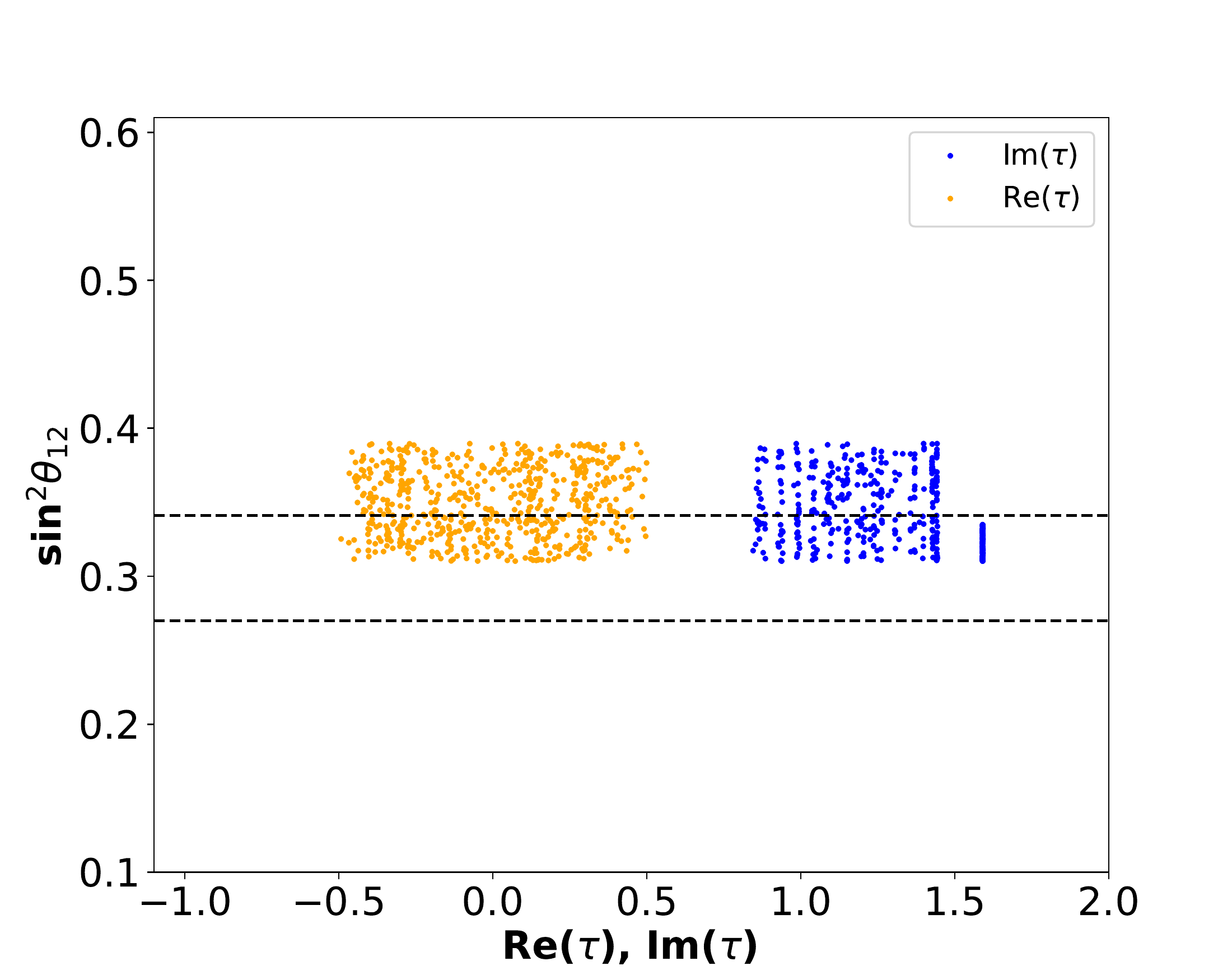}\label{rel-2}}
\hspace*{0.2 true cm}\\
\subfloat[]{\includegraphics[width=75mm, height=50mm]{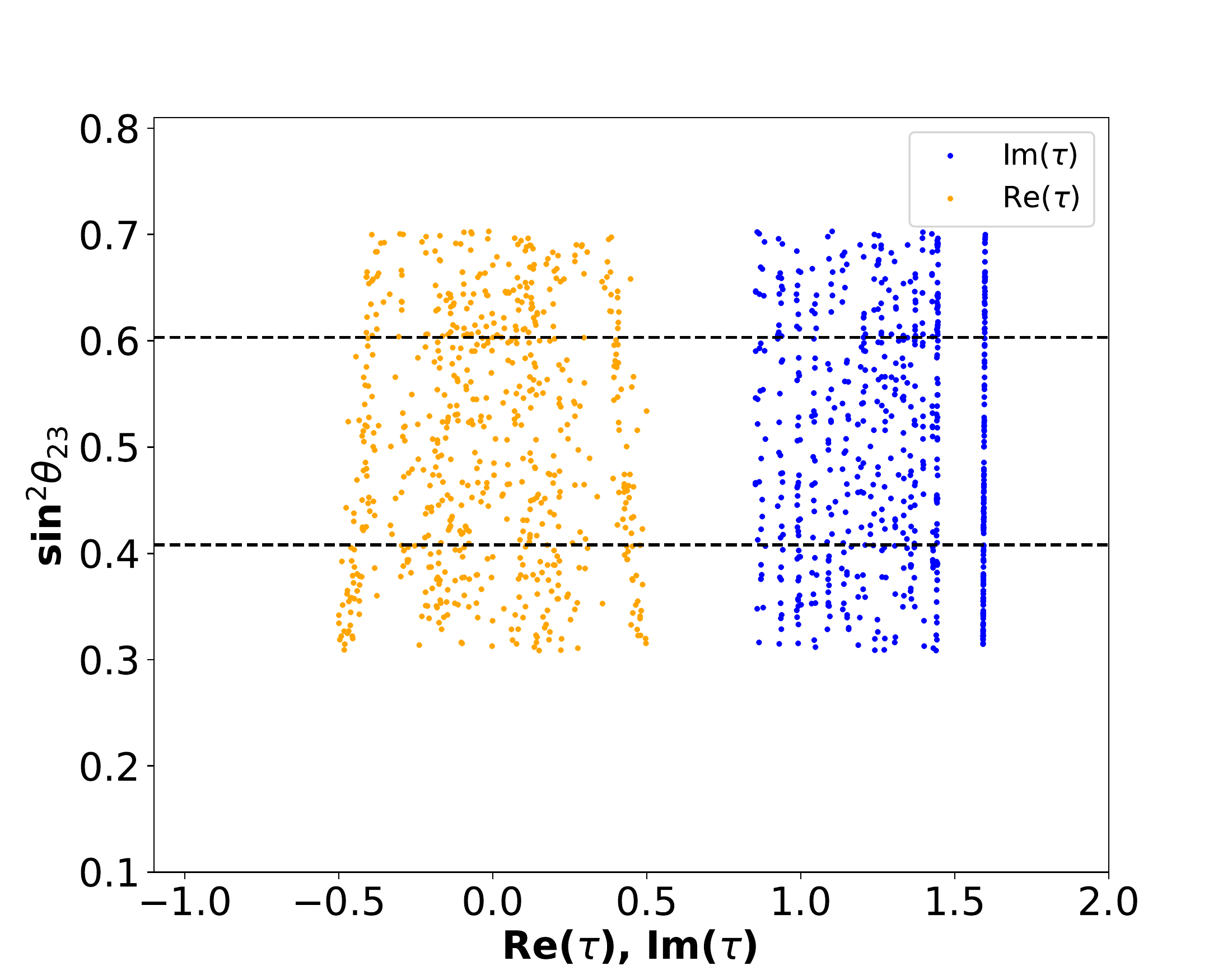}\label{rel-3}}

    \caption{In the above plots \ref{rel-1}, \ref{rel-2}, \ref{rel-3} depict the correlation of Re($\tau$) and Im($\tau$) with mixing angles $\sin^2 \theta_{13}$, $\sin^2 \theta_{12}$ and $\sin^2 \theta_{23}$ respectively. }
    \label{model-oscill}
\end{figure}

As a result, the left panel of Fig.(\ref{correl_mix_sum}) projects the interdependence between $\sin^2\theta_{13}$ (i.e. varying within [$0.02052-0.02398$]) w.r.t. the sum of neutrino masses ($\sum m_{\nu_i}$), where, the value of $\sum m_{\nu_i}$ is found to be above its lower bound, i.e., $0.058$ eV \cite{RoyChoudhury:2019hls, Planck:2018vyg} obtained for NH and assuming the lightest neutrino mass to be quite small. The right panel of Fig.(\ref{correl_mix_sum}) shows the interdependence of $\sum m_{\nu_i}$ with $\sin^2 \theta_{12}~(\sin^2 \theta_{23})$, where, it is seen that $\sin^2 \theta_{12}$  satisfies a very narrow region of [$0.311-0.341$] and  $\sin^2 \theta_{23}$ is within the range [$0.408-0.603$]. Further in Fig.(\ref{obs}), the panel (\ref{dcp-ss13}) shows the interdependence of $\sin^2\theta_{13}$ with CP phase  $\delta_{CP}$ which varies within $[142.1^\circ - 283^\circ]$, whereas the panel (\ref{M1-M2}) expresses the correlation of $M_{R_1}$ and $M_{R_2}$, i.e., heavy fermion masses and is found to be hierarchical, where, $M_{R_1}$ lies between [$0.5-13.4$] TeV, the lower limit  obtained for $M_{R_2}$ is  $128.8$ TeV going upto $5530$ TeV and finally in panel (\ref{jcp-ss13}) we depict the correlation of reactor mixing angle with Jarlskog invariant and see that $|J_{CP}| \leq 0.01$ with $\sin^2 \theta_{13}$ within its $3\sigma$ range. Proceeding further, in Fig.(\ref{model-oscill}), we depict the correlation of Re($\tau$) and Im($\tau$) with mixing angles (i.e., $\sin^2 \theta_{13}$~ [\ref{rel-1}], $\sin^2 \theta_{12}$~ [\ref{rel-2}] and $\sin^2 \theta_{23}$~ [\ref{rel-3}]) due to the fact that there is an implicit relation of oscillation parameters with modulus $\tau$.


\section{$W$ mass anomaly}
\label{sec:wboson}
The $W$ mass anomaly,  associated with the recent measurement of its value by the CDF-II collaboration \cite{CDF:2022hxs}, indicates the role of  physics beyond the standard model (BSM). Considering this discrepancy is just a consequence of the BSM, we  assume the mass of $W$ boson gets an immediate effect in the presence of the scalar super-partner of triplet superfield, i.e., $(\Tilde{\Sigma}_i)$ whereas, the mass of $Z$ boson remains unchanged \cite{Batra:2022org, Ghoshal:2022vzo}. Due to the hierarchical nature of fermion triplets as shown in the upper right panel of Fig.(\ref{obs}), it is assumed that the mass of scalar triplets $(\Tilde{\Sigma}_j)$ is also hierarchical. So, the VEV of the smallest scalar field will contribute positively to explain updated $W$ mass by CDF-II. The soft breaking terms in the presence of $\Tilde{\Sigma}_1$, in addition with the MSSM soft breaking term are \cite{Martin:1997ns, DiChiara:2008rg} given below,
\begin{eqnarray}
 -\mathcal{L}
                   & =&   m^2_{\scriptscriptstyle H_u} |H_u|^2 + m^2_{\scriptscriptstyle H_d} |H_d|^2 + b H_u H_d  +  2a_{\Sigma}^2 \lambda_1 {\rm Tr}(\Tilde{\Sigma}_{1} \Tilde{\Sigma}_1) + 2\lambda_2 B_{\lambda}\left( H_u^T \eta \Tilde{\Sigma}_1 H_d \right),
\end{eqnarray}
where, $m^2_{\scriptscriptstyle H_u}$, $m^2_{\scriptscriptstyle H_d}$, $b$, $a_{\Sigma}^2$ and $B_{\lambda}$ are soft breaking parameters, and   $\lambda_1$($\lambda_2$) have modular form with $k_I = 3(6)$ and transform as doublet under $T^\prime$ symmetry as defined in eqn.(\ref{A6}) and eqn.(\ref{lambda-couplings}) in Appendix \ref{app:A}. 
The scalar potential at  the tree level can be written as
\begin{eqnarray}
V &=&   (m^2_{\scriptscriptstyle H_u} + \mu ^2 ) |H^0_u|^2 +   (m^2_{\scriptscriptstyle H_d} + \mu ^2 ) |H^0_d|^2 +  \lambda_1(a_{\Sigma}^2  +  \lambda_1 M_{\Tilde{\Sigma}}^2)|\Tilde{\Sigma}_1^0|^2  -bH^0_uH^0_d \nonumber \\
&+& (B_\lambda  - 
   2 \lambda_1 M_{\Tilde{\Sigma}})\lambda_2(H_u^0\Tilde{\Sigma}_1^0H_d^0)  +  \lambda_2^2|\Tilde{\Sigma}_1^0|^2   (|H_u^0|^2+|H_d^0|^2) + 2 \mu \lambda_2  \Tilde{\Sigma}_1^0   (|H_u^0|^2+|H_d^0|^2)\nonumber \\  &+& \lambda_2^2|H_d
^0|^2|H_u^0|^2 
+ \frac{1}{8}(g_1^2+ g_2^2)(|H_u^0|^2-|H_d^0|^2)^2.
\label{poten}
\end{eqnarray}
 
\begin{figure}[htpb]
\begin{center}
\hspace*{0.2 true cm}
\includegraphics[height=50mm,width=75mm]{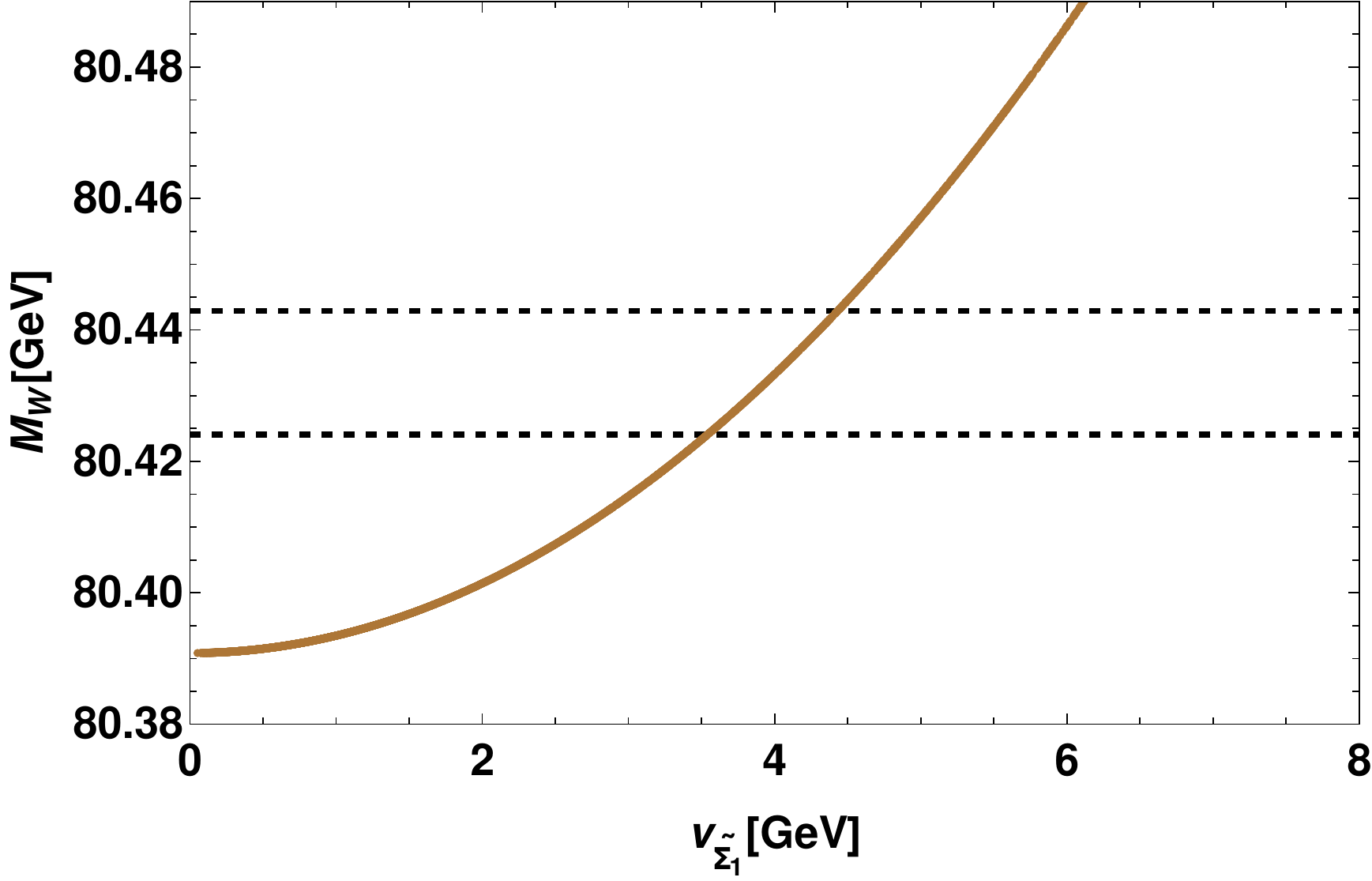}
\includegraphics[height=50mm,width=75mm]{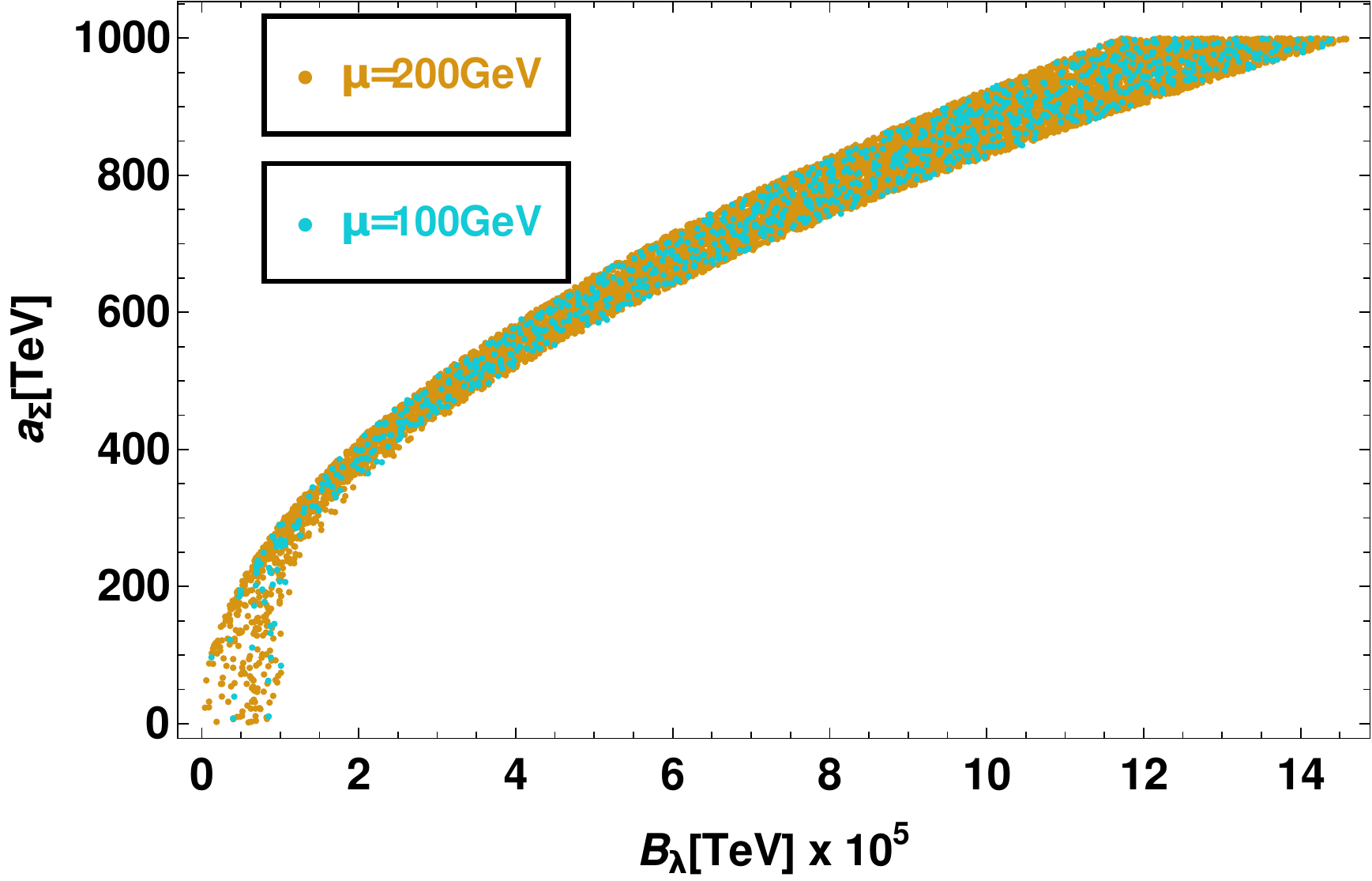}
\vspace{1cm}\\
\includegraphics[height=50mm,width=75mm]{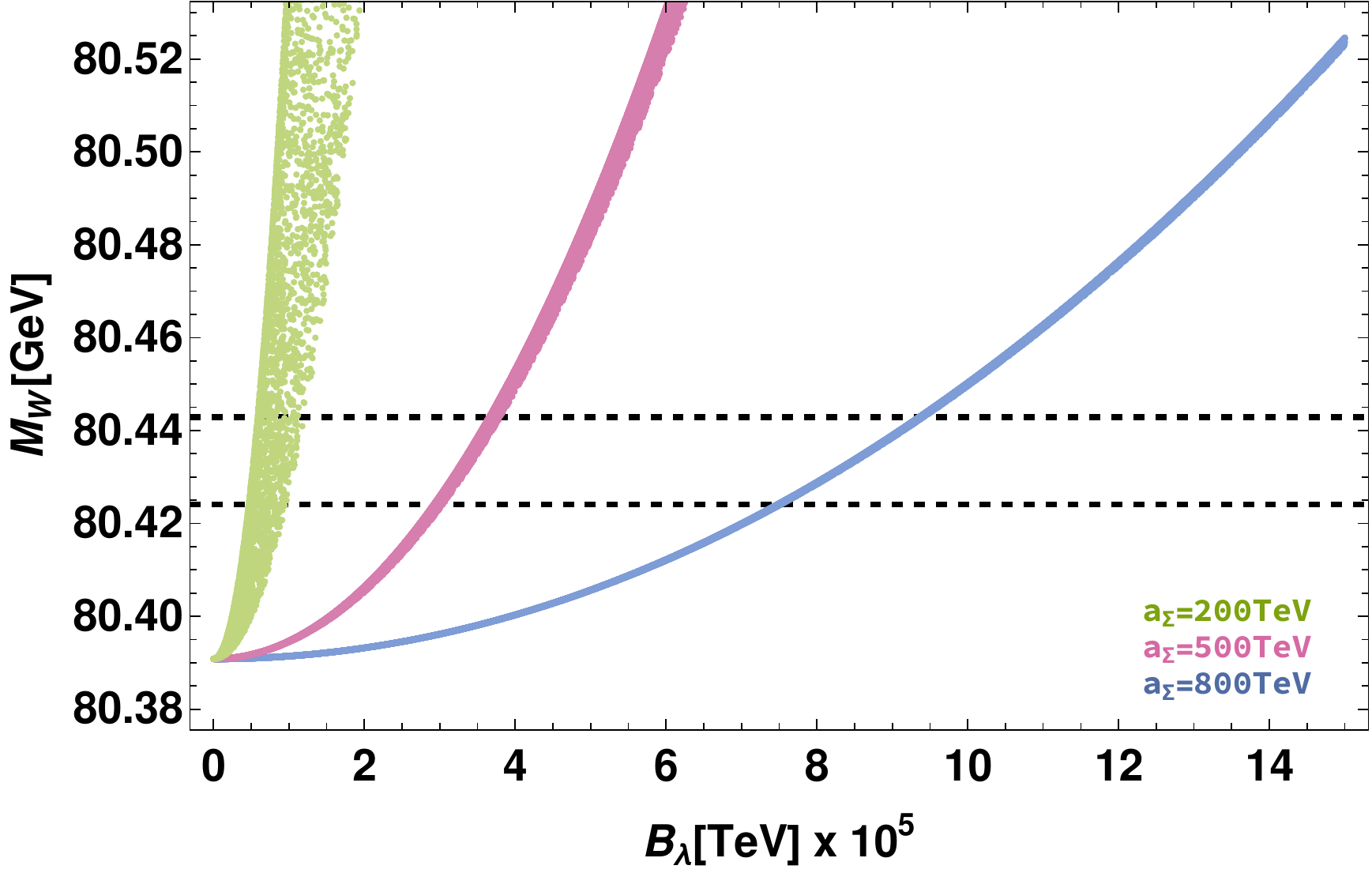}
\includegraphics[height=50mm,width=75mm]{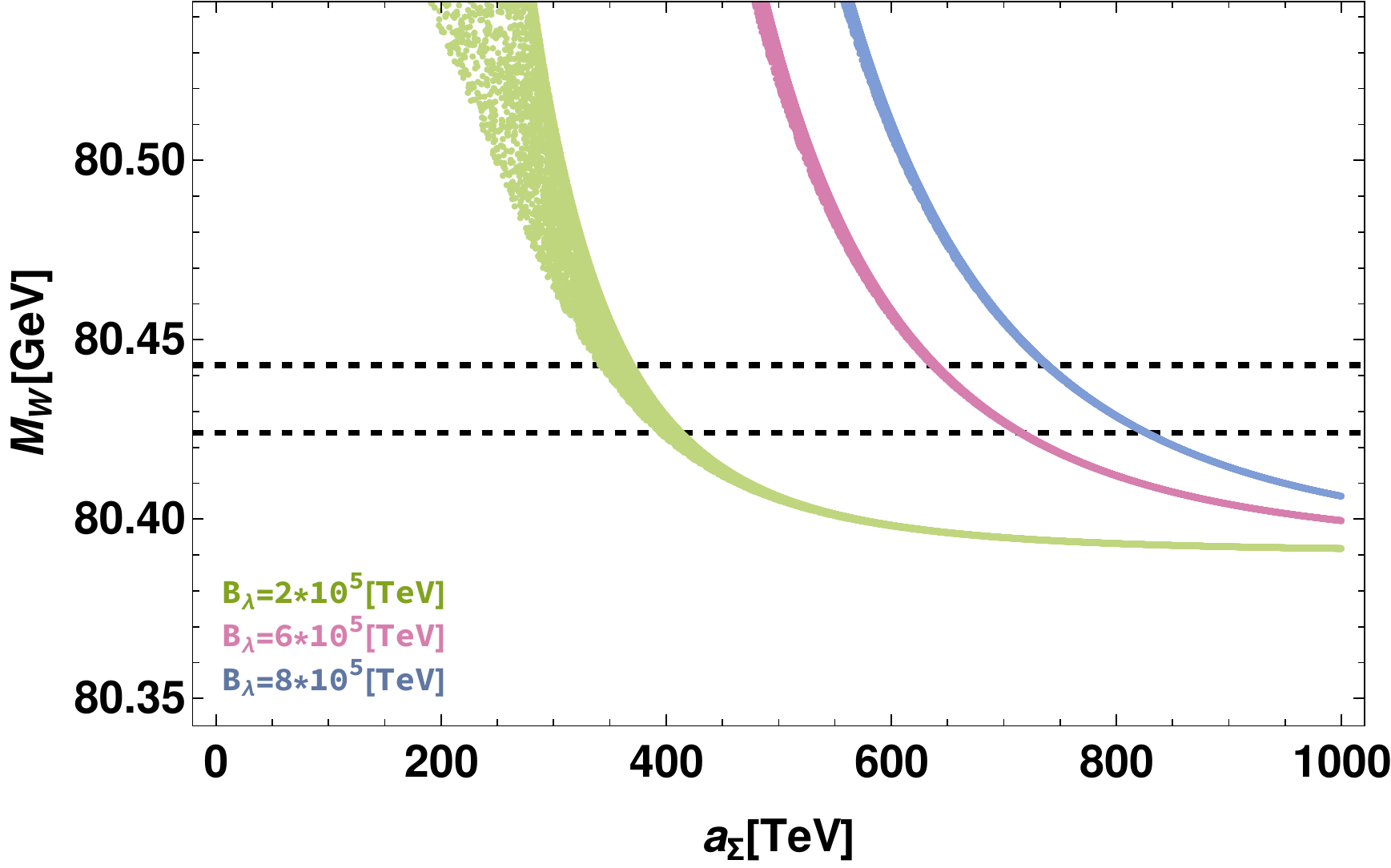}
\caption{The plot in the upper left panel displays the permissible range of vacuum expectation value (VEV) for the scalar triplet $\Tilde{\Sigma}_1$, which can elucidate the anomaly in the W boson mass. The upper right plot demonstrates the interdependence between $B_{\lambda}$ and $a_{\Sigma}$ when restricted to the 3$\sigma$ constraint of the W boson mass.  In the lower left (right) plot, the behavior of $B_{\lambda}$ ($a_{\Sigma}$) with respect to $M_W$ is presented for three distinct values, each represented by a different color. In all of these plots, the value of $\mu$ has been held constant at 150 GeV.}
\label{Msigma}
\end{center}
\end{figure}

The terms containing $\mu ^2$ as its coefficient come from F-term, whereas $g_1$ and $g_2$ are gauge couplings, resulting from D-term contribution to the scalar potential \cite{Martin:1997ns}.
After minimising
the scalar potential, we get the following conditions, which are utilized in the calculations of the mass of real part of Higgs, as elaborated in subsec. (\ref{mh}), 
\begin{eqnarray}
 m^2_{\scriptscriptstyle H_u} &=&  \frac{b\cot{\beta}}{2} - \mu ^2 -\frac{\lambda_2} {2\sqrt{2}} (B_{\lambda} -2  \lambda_1M_{\Tilde{\Sigma}})v_{\scriptscriptstyle \Tilde{\Sigma}_1^0} \cot{\beta} -\frac{\lambda^2_2}{2}( v^2_{\scriptscriptstyle \Tilde{\Sigma}_1^0}+v^2_d ) \nonumber \\ &-& \sqrt{2}\mu \lambda_2 v_{\scriptscriptstyle \Tilde{\Sigma}_1^0} -  \frac{1}{8}(g_1^2+g_2^2)(v_u^2 - v_d^2) ,\nonumber \\
 \label{mhu}
 m^2_{\scriptscriptstyle H_d} &=& \frac{b\tan{\beta}}{2} - \mu ^2 -\frac{\lambda_2} {2\sqrt{2}} (B_{\lambda} -2  \lambda_1M_{\Tilde{\Sigma}})v_{\scriptscriptstyle \Tilde{\Sigma}_1^0} \tan{\beta} -\frac{\lambda^2_2}{2}( v^2_{\scriptscriptstyle \Tilde{\Sigma}_1^0}+v^2_u ) \nonumber \\ &-& \sqrt{2}\mu \lambda_2 v_{\scriptscriptstyle \Tilde{\Sigma}_1^0} +  \frac{1}{8}(g_1^2+g_2^2)(v_u^2 - v_d^2) .
 \label{mhd}
\end{eqnarray}
The VEV of $\Tilde{\Sigma}^0
_1$ can be written as,
\begin{eqnarray}
v_{\scriptscriptstyle \Tilde{\Sigma}_1^0} &=& \frac{\lambda_2}{\lambda_1\sqrt{2}}\left( \frac{(\lambda_1M_{\Tilde{\Sigma}} -\frac{B_\lambda}{2}) v_u v_d-\mu v_{\scriptscriptstyle H}^2}{\lambda_1M_{\Tilde{\Sigma}}^2  + a_{\Sigma}^2+ \frac{\lambda_2^2}{2} v_{\scriptscriptstyle H}^2}\right),
\label{vsig}
\end{eqnarray}\\
which ultimately contributes only to the mass of $W$ boson, while $Z$ mass remains unchanged, as depicted below, 
\begin{eqnarray}
    M_{\scriptscriptstyle W}^2&=&\frac{1}{4}g_2^2 (v_{\scriptscriptstyle H}^2 +v^2_{\scriptscriptstyle \Tilde{\Sigma}_1^0}),~~
    M_{\scriptscriptstyle Z}^2 = \frac{v_{\scriptscriptstyle H}^2(g_1^2  + g_2^2 )}{4} .
    \label{MW2}
\end{eqnarray}
 We scan the assumed parameters in the following ranges \cite{DiChiara:2008rg}:
\begin{eqnarray}
  &&  \mu = [100,200] ~{\rm GeV } , ~~B_{\lambda} = [1,2\times 10^6]~{\rm TeV},~~ a_\Sigma = [1,10^3]~{\rm TeV}, \nonumber \\ &&~~~~~~~~M_{\Tilde{\Sigma}}=[10, 100]~{\rm TeV}, ~~ b=[10^2, 10^4]~{\rm TeV^2} \;.   
    \label{range2}
\end{eqnarray}
 In order to account for the new CDF-II result for the $W$ boson mass, the VEV of $\Tilde{\Sigma }^0_1$ must lie within a specific range. This range is identified as $3.5$ \rm{GeV} to $4.4$ \rm{GeV} and is shown in upper left panel of Fig.(\ref{Msigma}), from the variation of  $M_W$  with the $v_{\scriptscriptstyle \Tilde{\Sigma}_1^0}$. Also, under the roof of SM, the $\rho$ parameter value is given as,
\begin{eqnarray}
    \rho_{\scriptscriptstyle SM} = 1.00038 \pm 0.00020\;,
\end{eqnarray}
and the updated values of $\rho$ parameter due to $W$ mass from the CDF-II result, 
\begin{eqnarray}
    \rho_{\scriptscriptstyle CDF} = \frac{M_{\scriptscriptstyle W}^{2}}{M_{\scriptscriptstyle Z}^2 \cos^2 \theta_{w}} = 1.00179.
    \label{rhop}
\end{eqnarray}
We can define the $\rho$ parameter in terms of VEV's of $\Tilde{\Sigma}^0$,  $H_u$ and $H_d$,
\begin{eqnarray}
    \rho= 1+ 8\frac{v^2_{\scriptscriptstyle \Tilde{\Sigma}_1^0}}{v_{H}^2}\;.
      \label{rho}
\end{eqnarray}
It is worth noting that eqns.(\ref{rhop}) and (\ref{rho}) provide the value of $v_{\scriptscriptstyle{\Tilde{\Sigma}_1^0}} \simeq$ 3.5 GeV, which falls within the specified range illustrated in the upper left panel of Fig.(\ref{Msigma}). We show the correlation between $B_{\lambda}$ and $a_{\Sigma}$, imposing the constraint of $3\sigma$ range of $W$ mass, for two specific values of $\mu = 100 ~\rm GeV, 200 ~\rm GeV$, and the result is shown in upper right panel of Fig.(\ref{Msigma}), which indicates that there is not much difference for both the values of $\mu$. Therefore, we adopt a benchmark value of $\mu$ = $150 ~\rm GeV$ to explore the dependence of other parameters on the mass of the $W$ boson. To achieve this, we employed three benchmark values of $a_{\Sigma}$, as 200 TeV, 500 TeV, and 800 TeV, to get a good correlation between $M_W$ and $B_{\lambda}$ as shown in lower left panel of Fig.(\ref{Msigma}). From this figure, it should be noted that, as the value of $a_\Sigma$ increases the allowed range of $B_\lambda$ also increases, which can also be inferred from the top-right panel of Fig.(\ref{Msigma}). Similar behaviour can also be noticed, if we consider three representative values for $B_{\lambda}$:   $2 \times 10^5 $ TeV, $6 \times 10^5 $ TeV, and $8 \times 10^5 $ TeV, to obtain a correlation between $M_W$ and $a_{\Sigma}$, as shown in the lower right panel of Fig. \ref{Msigma}. 


\subsection{Mass of CP even Higgs}
\label{mh}
\begin{figure}[htpb]
\begin{center}
\centering 
\includegraphics[height=50mm,width=70mm]{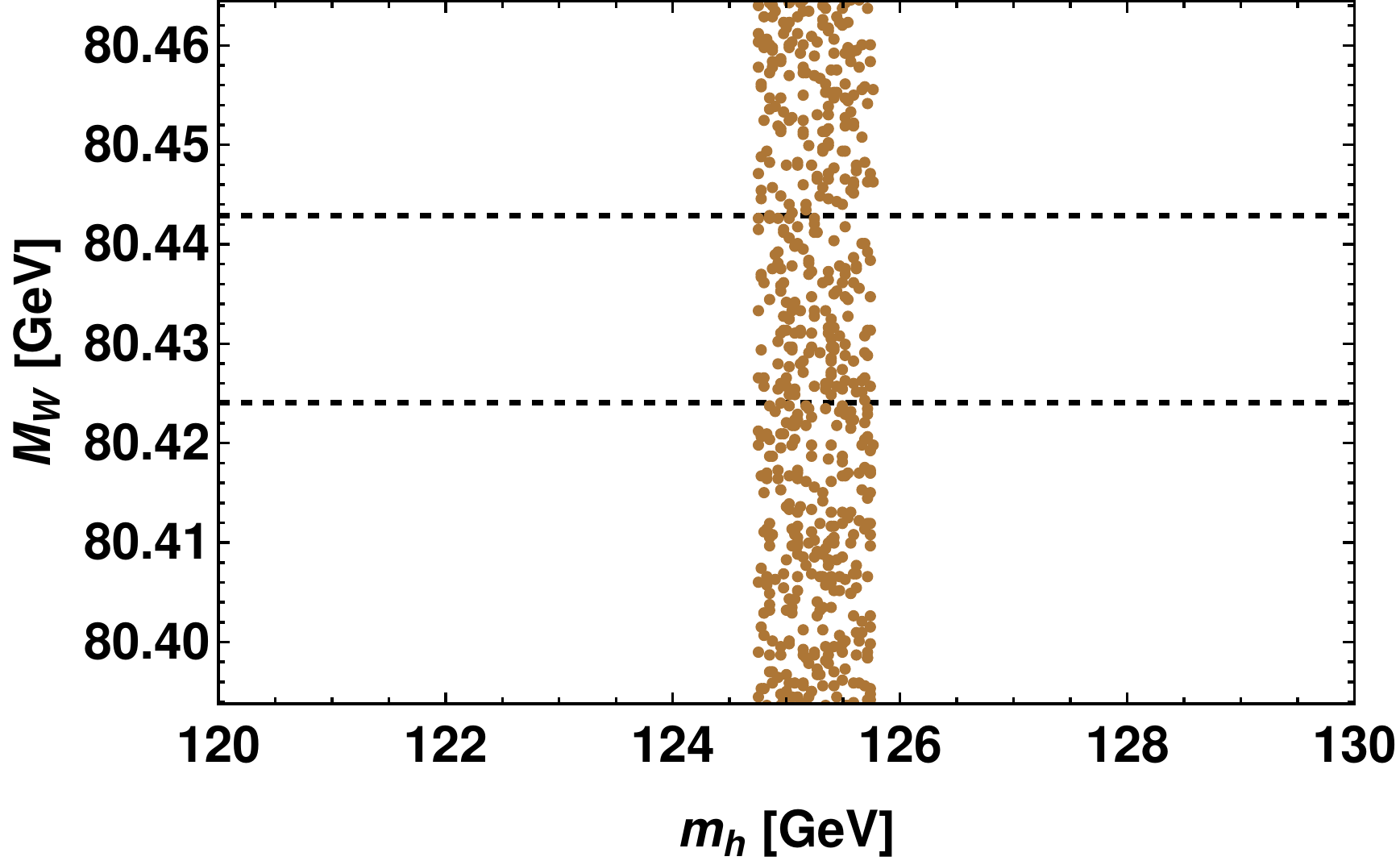}\\
\vspace{1cm}
\includegraphics[height=50mm,width=70mm]{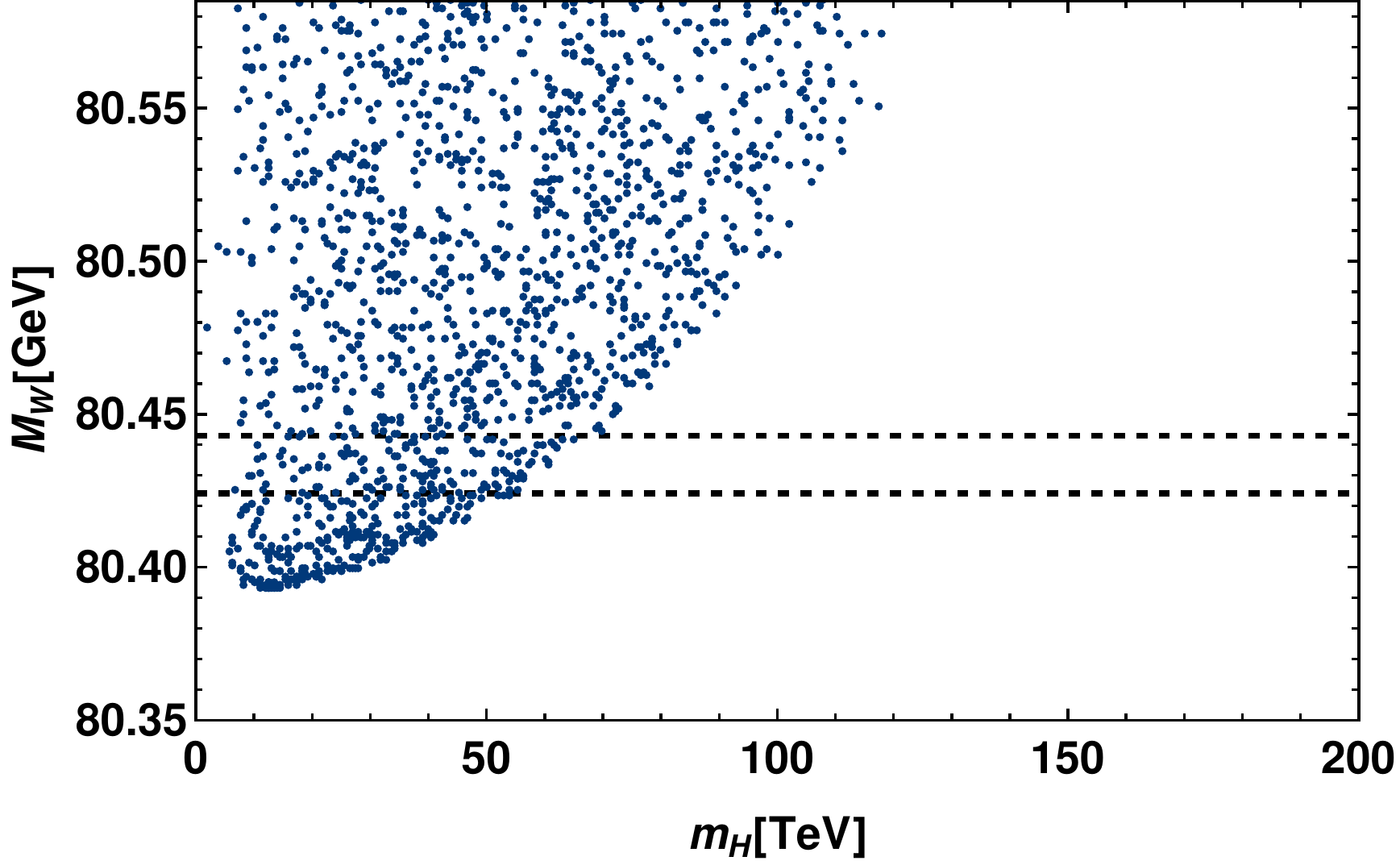}
\includegraphics[height=50mm,width=70mm]{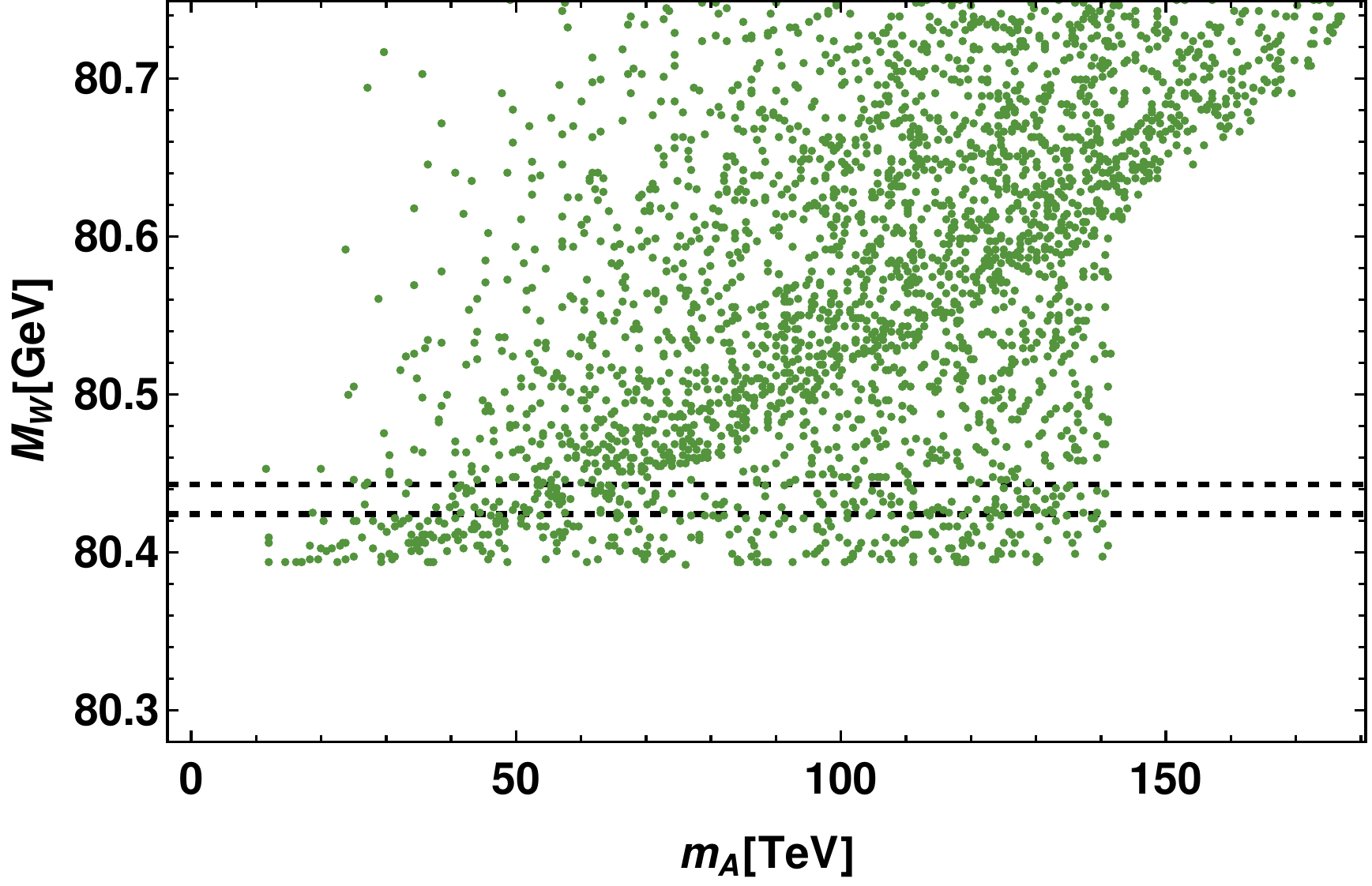}
\caption{The top panel displays a limit on the mass of the smallest scalar particle that has been obtained by imposing the mass of the $W$ boson.  The lower left (right) panel shows the limit bounded on BSM scalars $m_H$ ($m_A$) through $W$ mass.  }
\label{scalar-mass}
\end{center}
\end{figure}

 The neutral components of scalars can be written in terms of the real and imaginary parts as follows,
\begin{eqnarray}
    H^0_u &=& \frac{(H_{uR} + v_u) + iH_{uI}}{\sqrt{2}}, \\
    H^0_d &=& \frac{(H_{dR} + v_d) + iH_{dI}}{\sqrt{2}}, \\
    \Tilde{\Sigma}^0_1&=& \frac{(t_R + v_{\Tilde{\Sigma}_1^0}) + it_{I}}{\sqrt{2}},
\end{eqnarray}
where, $H_{uR}$, $H_{dR}$, $t_R$ are real and $H_{uI}$, $H_{dI}$, $t_I$ are imaginary parts of fields $H^0_u$, $H^0_d$, $\Tilde{\Sigma^0_1}$ respectively.
After electroweak symmetry breaking the symmetric mass matrix for CP even Higgs can be written in the basis of $(H_{uR},H_{dR}, t_R)$,
\begin{eqnarray}
    M^2_{\tiny{\scriptscriptstyle CP-even}} = \begin{pmatrix}
        m^2_{\scriptscriptstyle 11} & m^2_{\scriptscriptstyle 12} & m^2_{\scriptscriptstyle 13} \\
        m^2_{\scriptscriptstyle 21} & m^2_{\scriptscriptstyle 22} & m^2_{\scriptscriptstyle 23} \\
        m^2_{\scriptscriptstyle 31} & m^2_{\scriptscriptstyle 32} & m^2_{\scriptscriptstyle 33}
        \label{cpeven}
    \end{pmatrix},
\end{eqnarray}
with the matrix elements $m^2_{ij}$ as:
\begin{eqnarray}
    m^2_{\scriptscriptstyle 11} &=&  m^2_{\scriptscriptstyle H_u} + \mu ^2 + \frac{\lambda^2_2}{2}(v^2_{\scriptscriptstyle \Tilde{\Sigma}_1^0}+v^2_d ) + \frac{1}{8}(g_1^2+g_2^2)(3v_u^2 - v_d^2) + \sqrt{2} \lambda_2 \mu v_{\scriptscriptstyle \Tilde{\Sigma}_1^0} ,\nonumber \\
    m^2_{\scriptscriptstyle 22} &=&  m^2_{\scriptscriptstyle H_d} + \mu ^2 + \frac{\lambda^2_2}{2}(v^2_{\scriptscriptstyle \Tilde{\Sigma}_1^0}+v^2_u ) + \frac{1}{8}(g_1^2+g_2^2)(3v_d^2 - v_u^2) + \sqrt{2} \lambda_2 \mu v_{\scriptscriptstyle \Tilde{\Sigma}_1^0} ,\nonumber \\
    m^2_{\scriptscriptstyle 33} &=& \lambda_1 ( \lambda_1 M_{\Tilde{\Sigma}}^2 + a_{\Sigma}^2 ) + \frac{1}{2}\lambda_2^2 v_{\scriptscriptstyle H}^2, \nonumber\\
     m^2_{\scriptscriptstyle 12} &=& \lambda_2^2v_u v_d - \frac{b}{2} - \frac{1}{4}(g_1^2+g_2^2)v_uv_d + \frac{\lambda_2}{2\sqrt{2}}v_{\scriptscriptstyle \Tilde{\Sigma}_1^0}(B_\lambda -2 \lambda_1M_{\Tilde{\Sigma}} ),\nonumber \\
     m^2_{\scriptscriptstyle 13} &=& \frac{\lambda_2}{2\sqrt{2}} v_{d}(B_{\lambda} -2\lambda_1M_{\Tilde{\Sigma}} ) + \lambda_2^2 v_{\scriptscriptstyle \Tilde{\Sigma}_1^0} v_u +\sqrt{2} \mu  \lambda_2 v_u ,\nonumber\\
     m^2_{\scriptscriptstyle 23} &=& \frac{\lambda_2}{2\sqrt{2}} v_{u}(B_{\lambda} -2\lambda_1M_{\Tilde{\Sigma}} ) + \lambda_2^2 v_{\scriptscriptstyle \Tilde{\Sigma}_1^0} v_d +\sqrt{2} \mu  \lambda_2 v_d \;.
\end{eqnarray}\\
Since $M^2_{\tiny{\scriptscriptstyle CP-even}}$ is a symmetric matrix, so we have $m^2_{ij}$ = $m^2_{ji}$ and the expressions for $m^2_{11}$ and $m^2_{22}$ can be simplified further by using eqn.(\ref{mhd}). Diagonalization of the matrix  $M^2_{\tiny{\scriptscriptstyle CP-even}}$ provides mass for the real part of Higgs in basis $(h, H, \rm A)$. Fig. \ref{scalar-mass} illustrates  the constraints obtained on the masses of these three scalars  using the current observation of $W$ mass.   From the figure, we obtain  limits on their masses as  $m_h\in [124.74 , 125.76]$ GeV, corresponding to the SM Higgs, while $m_H \in [6.6 , 65.7]$ TeV and $m_A \in [18.7,140.8]$ TeV.

\section{Muon $(\lowercase{g-2})$}
\label{sec:g-2}
The triumph of the quantum field theory brings muon anomalous magnetic moment $(g-2)$ into the limelight. The convincing difference between measurements and predictions of the Standard Model (SM) could also portend new physics since it has historically drawn much attention. The SM contribution quantified so far is given as \cite{Aoyama:2012wk,Aoyama:2019ryr,Czarnecki:2002nt,Gnendiger:2013pva,Davier:2017zfy,Keshavarzi:2018mgv,Colangelo:2018mtw,Hoferichter:2019mqg,Davier:2019can,Keshavarzi:2019abf,Kurz:2014wya,Melnikov:2003xd,Masjuan:2017tvw,Colangelo:2017fiz,Hoferichter:2018kwz,Gerardin:2019vio,Bijnens:2019ghy,Colangelo:2019uex,Blum:2019ugy,Colangelo:2014qya},
\begin{align}
(a_{\mu})^{\rm SM}= 116 591 810(43)\times 10^{-11}.
\end{align}

\begin{figure}[htpb]
\centering 
\includegraphics[scale=0.5]{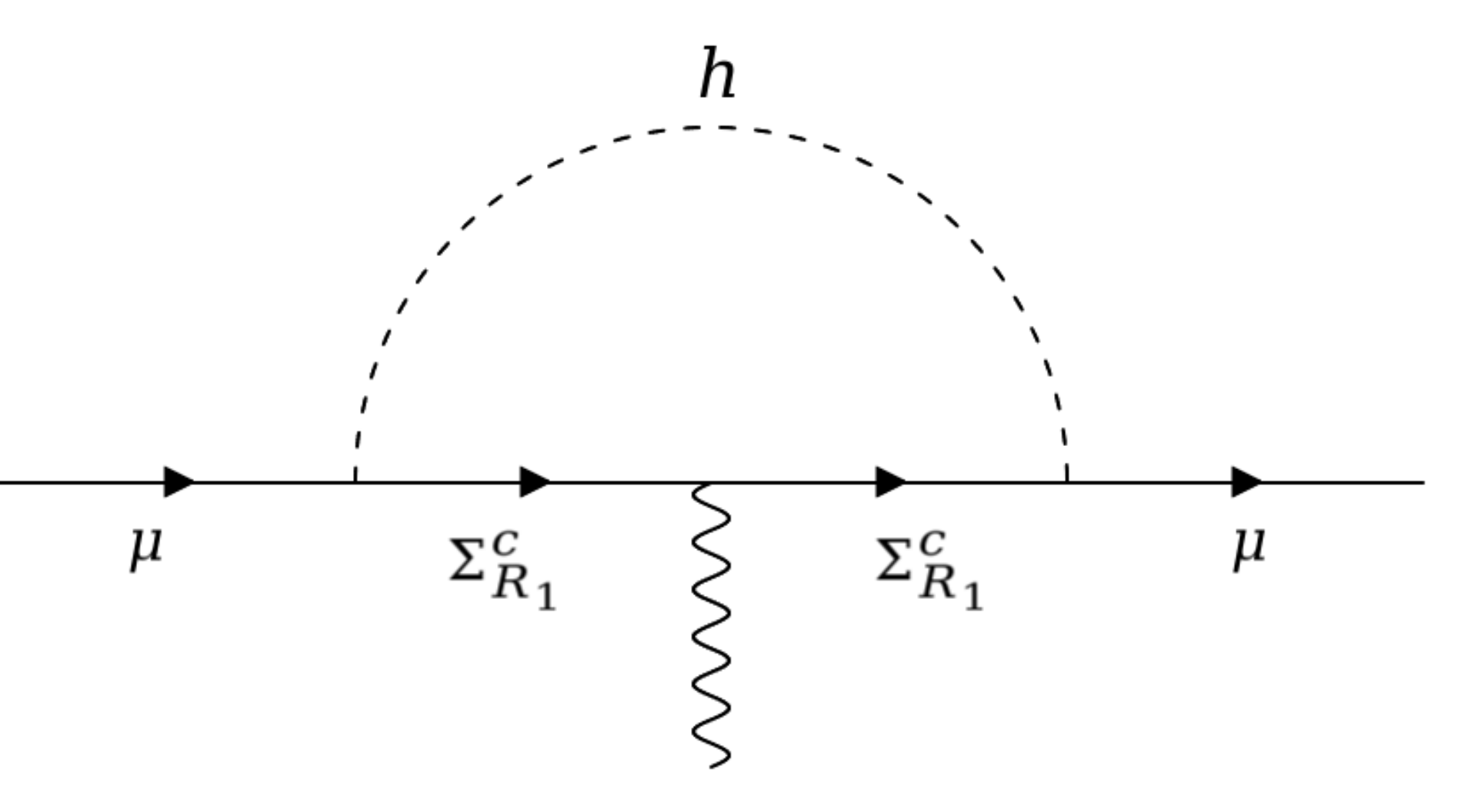}
\caption{Feynman diagram involving additional fermion triplet $\Sigma^c_{R_1}$ that generates muon anomalous magnetic moment.}
\label{feyn}
\end{figure}

As part of its April 2021 announcement,  Fermilab reported its first measurement on the muon anomalous magnetic dipole moment  \cite{Muong-2:2021vma} given below,
\begin{align}
(a_{\mu})^{\rm FNAL}= 116 592 040(54)\times 10^{-11} ,
\end{align} which  contradicts SM results by 3.3$\sigma$ and simultaneously agrees with the BNL E821 results \cite{Muong-2:2006rrc,Muong-2:2021ojo},
\begin{align}
(a_{\mu})^{\rm BNL}= 11659208.0(6.3)\times 10^{-10}.
\end{align} 
The size of the difference between the average of both experiments and SM prediction is,
\begin{align}
\Delta a_{\mu}= (a_{\mu})^{\rm exp}-(a_{\mu})^{\rm SM}= (251\pm 59)\times 10^{-11},~~~~~~~~~~~~~~
\end{align}
at 4.2$\sigma$ level. This deviation is significantly large enough, pointing towards the possible role of new physics. In this context, we show the   new  fermionic triplet $\Sigma^c_{R_1}$ could be a potential candidate for explaining  $(g-2)_{\mu}$ discrepancy. The relevant contribution is shown in Fig.(\ref{feyn}), obtained from the  corresponding superpotential term, i.e., the second term in Eq. (\ref{comp_superpotential}).

\begin{figure}[htpb]
\begin{center}
\centering 
\includegraphics[height=55mm,width=75mm]{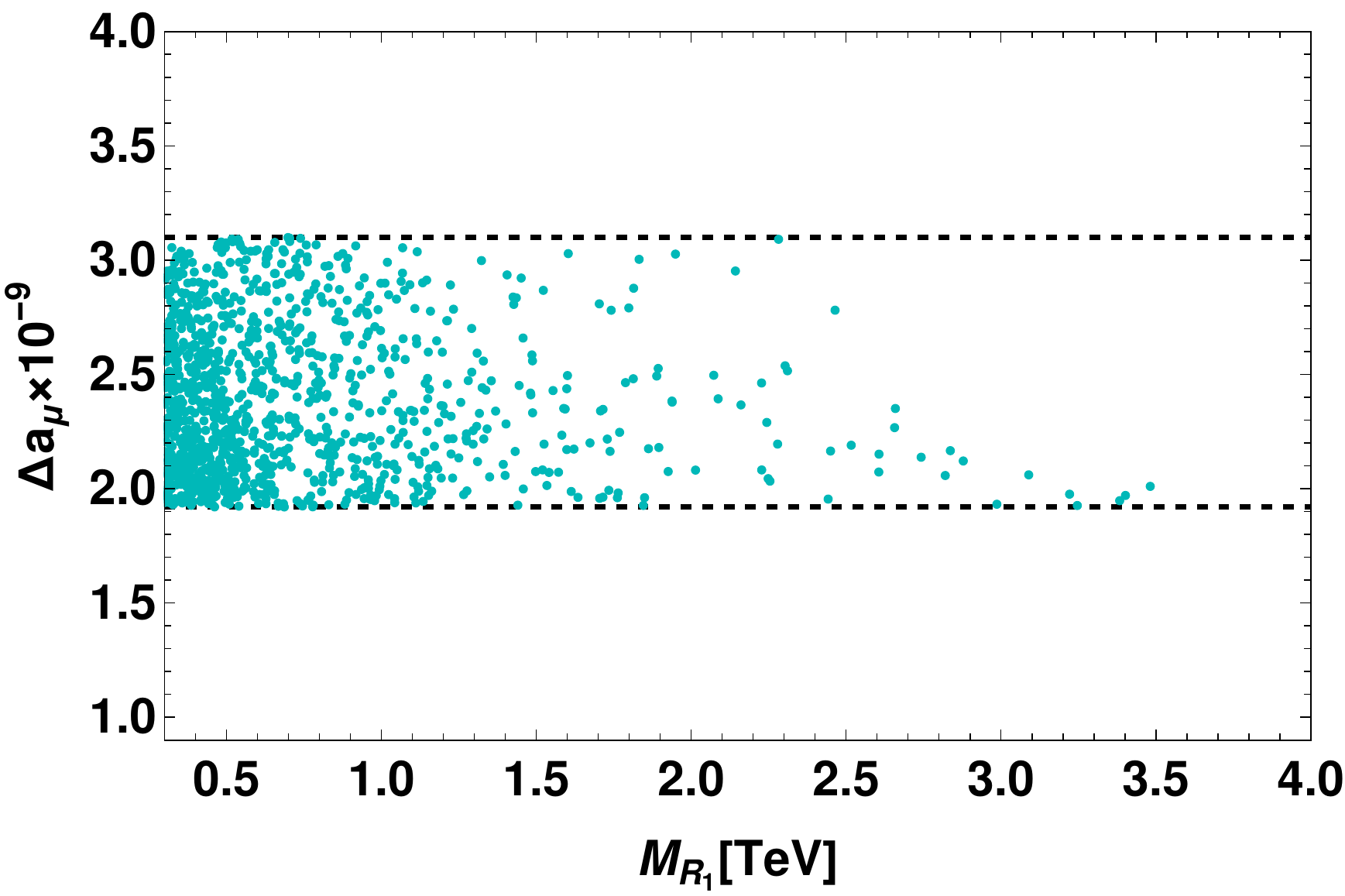}
\caption{The above  panel shows the contribution of the lightest fermion triplet to muon $(g-2)$.   }
\label{Fig:g-2}
\end{center}
\label{even-scalar}
\end{figure}

Thus, we obtain the additional contribution to muon $(g-2)$ as \cite{Kannike:2011ng},
\begin{align}
\Delta a_{\mu} = & 
  \frac{m_{\mu}^2}{32\pi^2 m_{h}^2} \{({2|g_{\scriptscriptstyle D_1} y_{22}|^2})F_h(x_1) + z_1  Re[(g_{\scriptscriptstyle D_1} y_{22})^2]G_h(x_1)\},
\end{align}
where, $x_1$ = $\frac{M_{R_1}^2}{m_h^2}$, $z_1$ = $\frac{M_{R_1}}{m_\mu}$ and $g_{\scriptscriptstyle D_1}$ is free parameter defined in eqn.(\ref{free-para}). The loop functions are expressed as
 \begin{eqnarray}
    F_h(x_1) &=& \frac{x_1^3-6x_1^2 +3x_1+2+6x_1 \ln(x_1)}{6(1-x_1)^4},\\
   G_h(x_1) &=& \frac{-x_1^2 + 4x_1-3-2 \ln(x_1)}{(1-x_1)^3}.
\end{eqnarray}
 As the right-handed triplets have hierarchical mass, hence, only the lightest heavy fermion $\Sigma^c_{R_1}$ contributes towards muon anomalous magnetic moment. The correlational  behaviour of the mass of $\Sigma^c_{R_1}$ w.r.t. $\Delta a_{\mu}$ for $m_h = $125.25 GeV  is shown in  Fig.(\ref{Fig:g-2}).
 
\begin{figure}
    \centering
    \includegraphics[width=100mm, height=75mm]{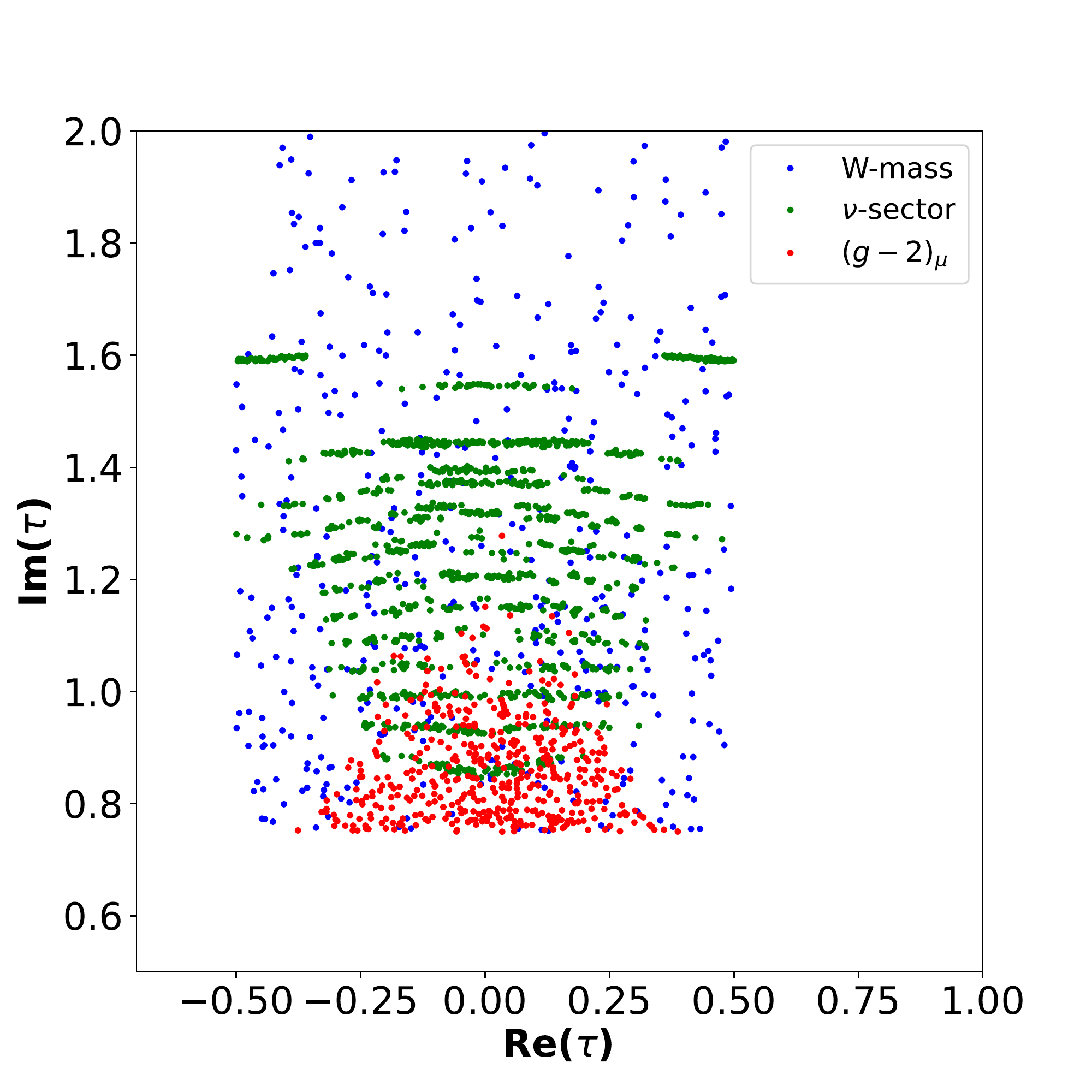}
    \caption{The points in blue (red) color satisfy $W$ mass ($(g-2)_\mu$) and green color data points are for neutrino phenomenology.}
    \label{fig:x-y}
\end{figure}

 Next, we would like to see the common allowed ranges on the values of Real and Imaginary parts of the modulus $\tau$ compatible with the neutrino oscillation phenomenology, $W$ mass, and muon $(g-2)$. In Fig. \ref{fig:x-y}, we present a plot illustrating the corresponding allowed parameter space compatible with  $W$ mass (represented by blue points), neutrino phenomenology shown by green points  and muon $(g-2)$ as depicted by red points.  From the figure,  we  obtain the ranges as $-0.27 \leq \rm Re(\tau) \leq 0.27$ and  $0.84 \leq \rm Im(\tau) \leq 1.15$, which satisfy all the three phenomenological aspects, discussed in this paper. 

\section{Lepton Flavor Violation}
\label{sec:LFVs}
In this section, our focus has been on exploring lepton flavor-violating decay as a means of establishing more precise limitations on the mass range of heavy neutrinos. Of particular interest is the highly acclaimed and rare $(\mu \rightarrow e \gamma)$ decay mode, which represents one of the most strictly restricted modes to date, with current limits set at $4.2 \times 10^{-13}$ \cite{MEG:2016leq}. This mode is characterized by the fact that it cannot occur at the tree level and is associated with a lepton number violation. The decay widths and branching ratios for different lepton flavor-violating decays within the type-III seesaw model are presented in \cite{Abada:2008ea}. The heavy neutrino contribution, i.e., $M_{R_1}$ to the one-loop branching ratio \cite{Abada:2008ea, Chang:1993kw} of $\mu \rightarrow e \gamma$ is given below, 
\begin{equation}
    {\rm Br}(\mu \to e \gamma) = \frac{3m_e \alpha}{4\pi m_{\mu}}\left |(g_{\scriptscriptstyle{D_1}} y_{22})(g_{\scriptscriptstyle{D_1}} y_{22^{\prime\prime}} ) \frac{M^2_{R_1}}{m_h^2}\left (\frac{3}{2} + \ln\frac{M^2_{R_1}}{m_h^2} \right ) \right |^2,
\end{equation}
with $\alpha$ being the fine structure constant, $g_{D_1}$ being the free parameter, $y_{22}$ and $y_{22^\prime}$ are modular Yukawa couplings mentioned in table \ref{tab:Yukawa} and $m_e, m_\mu, m_h$ are the mass of the electron, muon, Higgs respectively.

\begin{figure}[htbp]
    \centering
\includegraphics[height=50mm,width=75mm]{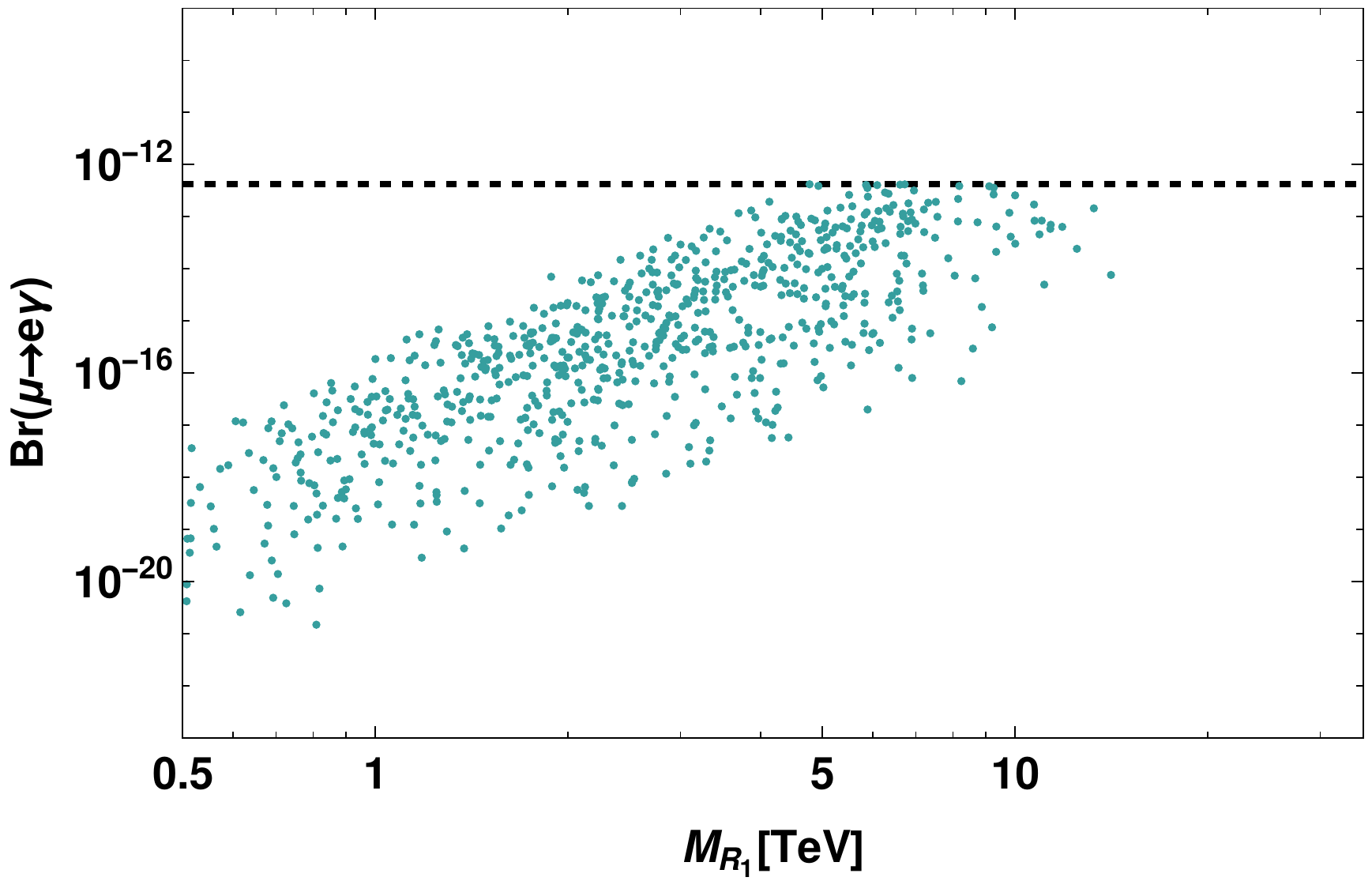}    
\caption{Variation of ${\rm Br} (\mu \to e \gamma)$ against $M_{R_1}$ (TeV), where the gridline  shows the experimental upper bound.}
    \label{LFV}
\end{figure}
The parameter space mentioned in sec. \ref{sec:numerical} are utilized to perform LFV, which mutually satisfies neutrino phenomenology and other phenomena discussed in our paper. The plot for the branching ratio of $(\mu \to e\gamma)$  is depicted in Fig.(\ref{LFV}) with respect to $M_{R_1}$,  where the black dashed horizontal line represents the experimental upper limit \cite{MEG:2016leq}. From the figure, we find  the upper limit on the $M_{R_1}$ as $13.4$ TeV, consistent with the LFV decay $\mu \to e \gamma$.  This observation underscores the importance of considering the LFV bounds when investigating or constraining the limit of the lightest heavy neutrino mass, i.e., $M_{R_1}$ in such models.

\section{Conclusion}
\label{sec:conclude}
To comprehend neutrino phenomenology and explain observed oscillation data, we have considered a model including $ A_4^\prime$ modular symmetry employing type-III seesaw mechanism in a minimal super-symmetric context, i.e., adding only two $SU(2)_L$ triplet fermions ($\Sigma^c_{R_j}$). This yields a specific mass structure for Dirac and Majorana terms which further yields a $3\times3$ active neutrino mass matrix. There are various modular Yukawa couplings involved in keeping the superpotential invariant under $T^\prime$ modular discrete symmetry
for the explanation of the recent W-mass anomaly, where acquisition of VEV by modulus $\tau$ breaks $A_4^\prime$ symmetry. Here, the numerical diagonalization technique lifts the workload in the analytical part, and the results are predicted following the 3$\sigma$ constraint established through numerous experiments.  Consequently, we obtain the sum of active neutrino masses $\sum m_{\nu_i}$ within $[0.058-0.12]$ eV, and mixing angles are seen to be within their respective 3$\sigma$ ranges. Proceeding further, the results for  $\delta_{\rm CP}$ and Jarlskog invariant $|J_{\rm CP}|$ are seen to be within $[142.1^\circ - 283^\circ]$ respectively, establishing a firm correlation. Further, from the upper bound on the  Br($\mu \to e\gamma$), the mass of the lightest right-handed neutrino is highly constrained; hence, the mass range for $M_{R_1}$ is found to be [$0.05-13.42$] TeV and that of $M_{R_2}$ is in the range of [$128.8-5530$] TeV, establishing a hierarchy between them. Advancing further, we attempt to explain the $W$ mass anomaly, where the presence of the scalar super-partner impacts the result, and the new mass range for $W$-mass is $80.4335 \pm 0.0094$ GeV. Finally, we were successful in accommodating the results from muon $(g-2)$ explaining the recent results.

\acknowledgments 
PM wants to thank Prime Minister's Research Fellowship (PMRF) scheme for its financial support. MKB would like to thank DST-Inspire for its financial support. RM would like to acknowledge University of Hyderabad IoE project grant no. RC1-20-012. We gratefully acknowledge the use of CMSD HPC facility of Univ. of Hyderabad to carry out the computational work.

\appendix
\section{T$^\prime$ Modular Symmetry}
\label{app:A}
\
The Modular forms of couplings required in our model are given as,

\begin{itemize}
    \item Modular forms transforming as doublet under $T^\prime$ symmetry and with modular weight $k=1$:
    \begin{eqnarray}
        Y_2^{(1)}(\tau) = \begin{pmatrix}
            Y_1 \\
            Y_2
        \end{pmatrix}
    \end{eqnarray}
    where, $Y_1$ and $Y_2$ are the function of $\tau$ and are defined as,

\begin{eqnarray}
Y_1 &=& \sqrt{2}e^{i 7\pi/12}q^{1/3}(1+q+2q^2+2q^4+q^5+2q^6+.....),\\ \nonumber
Y_2 &=& 1/3+2q+2q^3+2q^4+4q^7+2q^9+.....,
\end{eqnarray}
with $q=e^{i2\pi \tau}$.
\item The modular forms for the Yukawa couplings required to write superpotential term for   neutral lepton sector are with modular weight 5:
\begin{eqnarray}
Y_{2,I}^{(5)}= \begin{pmatrix}2\sqrt{2}e^{i 7\pi/12} Y_1^4Y_2 + e^{i \pi/3} Y_1Y_2^4 \\ 2\sqrt{2}e^{i 7\pi/12} Y_1^3Y_2^2+e^{i \pi/3} Y_2^5
\end{pmatrix},~~~
\end{eqnarray} 
\begin{eqnarray}
Y_{2^\prime,I}^{(5)}= \begin{pmatrix}-Y_1^5 +  2(1-i)Y_1^2Y_2^3 \\ -Y_1^4Y_2+2(1-i)Y_1Y_2^4
\end{pmatrix},~~~
\end{eqnarray}
\begin{eqnarray}
Y_{2^{\prime\prime}}^{(5)}= \begin{pmatrix}5e^{i \pi/6} Y_1^3Y_2^2-(1-i)e^{i \pi/6}Y_2^5  \\ -\sqrt{2}e^{i 5\pi/12} Y_1^5-5e^{i \pi/6} Y_1^2Y_2^3
\end{pmatrix}.
\end{eqnarray} 

\item Couplings $\lambda_1$ and $\lambda_2$ have the forms:

\begin{eqnarray}
\lambda_1 = Y_{3,I}^{(6)} = \begin{pmatrix}
-2(1-i)Y_1^3Y_2^3+iY_2^6\\ -4e^{i \pi/6} Y_1^4Y_2^2-(1-i)e^{i \pi/6} Y_1Y_2^5 \\ 2\sqrt{2}e^{i 7\pi/12} Y_1^5Y_2+e^{i \pi/3} Y_1^2Y_2^4
\end{pmatrix},~~~
\label{A6}
\end{eqnarray}
\begin{eqnarray}
\lambda_2 = Y_{2^{\prime\prime}}^{(3)} = \begin{pmatrix}Y_1^3 + (1-i)Y_2^3 \\ -3Y_2 Y_1^2
\end{pmatrix}.
\label{lambda-couplings}
\end{eqnarray}
\end{itemize}
\bibliographystyle{my-JHEP}
\bibliography{bib}
\end{document}